\documentclass[10pt, conference, twocolumn,a4paper]{IEEEtran}  

\usepackage{setspace}
\usepackage{colortbl}
\usepackage{amsmath}
\usepackage{amsthm}
\usepackage{amssymb}
\usepackage{verbatim}
\usepackage{bm}

\newcommand{\source}{  }

\newcommand{\bss}{ {\bm{\bm s}} }

\newcommand{\bmell}{ {\bm{\ell}} }

\newcolumntype{R}[1]{>{\raggedleft\arraybackslash }b{#1}}
\newcolumntype{L}[1]{>{\raggedright\arraybackslash }b{#1}}
\newcolumntype{C}[1]{>{\centering\arraybackslash }b{#1}}

\usepackage{graphicx}
\usepackage{epstopdf}
\usepackage{setspace}

\usepackage{tikz}
\usepackage{subfigure}
\usepackage{epsfig}

\usepackage{url}

\usepackage{tikz}
\usepackage{verbatim}
\usetikzlibrary{arrows,shapes}

\usetikzlibrary{shapes,snakes}
\usetikzlibrary{positioning,chains,calc}
\usepackage{relsize}
\usetikzlibrary{patterns}
\usepackage{colortbl}
\definecolor{kugray5}{RGB}{224,224,224}

\usepackage[linesnumbered,ruled,vlined]{algorithm2e}

\usepackage{pgfplots}
\pgfplotsset{compat=newest}

\usepgfplotslibrary{patchplots}

\theoremstyle{plain}
\newtheorem{lemma}{Lemma}
\newtheorem{theorem}{Theorem}
\newtheorem{proposition}[theorem]{Proposition}

\theoremstyle{definition}

\allowdisplaybreaks[4]
\IEEEoverridecommandlockouts

\begin{document}

\newlength{\figurewidth}\setlength{\figurewidth}{0.6\columnwidth}



\title{Incentive-rewarding  mechanisms to stimulate participation in heterogeneous DTNs}
\author{Rachid  El-Azouzi$^\star$,  Ahmed El Ouadrhiri$^\circ$,  Balakrishna Prabhu$^\dagger$,  Daniel Menasch\'e$^\ddagger$ and Olivier Brun$^\dagger$
\thanks{
$^\star$CERI/LIA, University of Avignon, 
Avignon, France. 
\newline  $^\circ$ LIMS, University of  Sidi Mohammed Ben Abdellah, Fez, Morocco. 
\newline $^\dagger$ LAAS-CNRS, Universit\'e de Toulouse, CNRS, Toulouse, France.
\newline $^\ddagger$  Federal University of Rio de Janeiro, Brazil
}}
\maketitle
\thispagestyle{empty}

\newcounter{one}
\setcounter{one}{1}
\newcounter{two}
\setcounter{two}{2}
\newcounter{three}
\setcounter{three}{3}
\newcounter{four}
\setcounter{four}{4}
\author{	 
\IEEEcompsocitemizethanks{




}
}
\IEEEtitleabstractindextext{%
\begin{abstract}
Delay Tolerant Networks (DTNs) rely on the cooperation  of nodes in a network to forward a message from its source to its destination.  Most of previous studies on DTNs have focused on the design of routing schemes  under 
the hypothesis that each relay node is willing to participate in the forwarding process. However, the delivery of a message incurs  energy and memory costs. In this paper we handle the problem of how to incentivize  mobile nodes to participate 
in relaying messages using a reward mechanism. We 
 consider  heterogeneous relay nodes where the cost for taking part in the forwarding process varies as a function of  the mobility patterns of the relays.
 We show that under fairly weak assumptions on the mobility pattern, the expected reward the source pays remains the same irrespective of 
the information it conveys to the relays, provided that the type of information does not vary dynamically over time.  We also characterize the effect of time to live (TTL) counters  on the delivery probability and memory usage.  Using simulations based on a 
synthetic mobility model and real mobility traces, we perform a few tests that show a good accordance among theoretical and simulation results. 
\end{abstract}
\begin{IEEEkeywords}
Delay tolerant networks, Reward incentive mechanism, Information setting. 
\end{IEEEkeywords}}

\maketitle
\IEEEdisplaynontitleabstractindextext

\section{Introduction} \label{sec:intro}
Delay Tolerant Networks (DTNs), also known as Intermittently Connected Mobile Networks, are mobile wireless networks in which there is no persistent end-to-end link from a source to a destination. 
DTNs take advantage of mobile nodes to ensure the communication between  users; the connectivity is maintained by mobile nodes which  communicate when they are in range of each other. When a node receives a message and  is not in range of any other node at that time, 
it stores the message in its buffer and forwards it to another node whenever a  useful communication opportunity arises.

Various routing protocols have been proposed for DTNs, their aim being to increase the delivery probability at low costs. Some routing protocols attempt to enhance the delivery ratio by aggressively  replicating the messages (e.g., flooding).  Other 
protocols  employ specially crafted  metrics, which capture some sort of knowledge of the domain under consideration, to achieve this purpose (e.g., forwarding routing)  \cite{Bogo}.



Two-hop routing  provides a good compromise between  delay and resource consumption when compared against flooding or epidemic routing~\cite{TwoHop}. In addition, from the standpoint of 
feasibility and implementation, the optimal control can be implemented by the source node, with no need to install new routing control protocols on relay nodes; one can   prescribe  target performance metrics (e.g., delay)   and then determine 
to which relays  the source node should deliver messages to, 
and for how long the delivery process should last~\cite{TwoHop, TwoHop2}.  Currently, there are no corresponding results for  multi-hop routing protocols. Finally, it has been shown in \cite{TwoHop2} that if we have a network with a complete mixing 
of the trajectories of the nodes, two-hop routes are sufficient to achieve the    throughput capacity region of the network. For this reason, we shall assume a two-hop  routing protocol in the remainder of this paper. 

\subsection{Problem Statement}
DTNs overcome the problems associated to  intermittent connectivity and guarantee  communication through cooperation between nodes; each node is supposed to be willing to participate in relaying messages of other nodes. However, as in any fully distributed system, nodes may misbehave; certain nodes may not be able  to relay messages of other nodes in order to save  resources (especially energy), which in turn degrades network performance. Hence,  one of the fundamental questions in the realm of DTNs concerns incentives: \emph{how to incentivize  mobile 
nodes to participate in relaying messages?}


To handle this problem, in \cite{TMC16} a reward mechanism was proposed in order to encourage nodes to relay messages of others.  Under this mechanism, the source  node promises a reward  to each relay who accepts relaying its message, but informs them 
that only the first one to deliver the message to the destination will effectively receive the reward.  
 A relay accepts to forward the message to the destination if the  proposed reward offsets its expected cost, as estimated by the relay when it meets the source.  
The assessment of the  reward expected  by the relay  depends on the time at which it meets the source, and on any information given by the source to the relay that may assist the relay in the estimation of  its probability of success. 
The model proposed in \cite{TMC16}  assumed a homogeneous mobility for all relays:  all the relays have the same statistics for the inter-contact time with the source and the destination. However, in many applications  real traces reveal 
the existence of heterogeneity in mobile relays \cite{Picu2015}. Indeed, in real traces~\cite{romaTaxi} we observed  that different types of  mobile nodes have very different characteristics in terms of their communication ability and mobility patterns. 


In this paper, we generalize the results of \cite{TMC16} to a heterogeneous mobility model in which relays can have different statistics for inter-contact times. The heterogeneity also gives rise to a richer set of structures of information sharing from which the source can choose from, and its analysis involves an application of results on order statistics~\cite{david1981order}. 
When the source meets a relay, it has now  four options of information sharing strategies, as opposed to the three discussed in \cite{TMC16}.  

1) \textbf{full information: } the source reveals to the relay the number and the age of all existing copies of the message in the network.

2) \textbf{partial information: } 
 in this case we distinguish two options:  $(a)$ the relay will  be informed about the identities of the relays that  got the message before or $(b)$ the relay will know the number of copies circulating in the network, but not the identities of nodes involved in previous contacts. 
	
	We note that in the homogeneous case $(a)$ and $(b)$ give the same information.  In the heterogeneous case considered in this paper, in contrast,  knowing the identity (and the mobility pattern) of nodes involved in previous contacts gives additional information for the computation of the success probability.
		
3) \textbf{no information: }  the source does not reveal any information to the relay, in which case the relay only knows at what time it met the source. 


\subsection{Related studies}

There is a vast literature on incentive schemes  to promote cooperation in DTNs \cite{ElAzouzi2012, giovani, chahin2013incentive,  TMC16, shevade2008incentive, chen2010mobicent, lu2010pi}.  Shevade \emph{et al.}~\cite{shevade2008incentive} uses Tit-for-Tat (TFT) to design an incentive-aware routing protocol that allows selfish DTN nodes to maximize their individual utilities while conforming to TFT constraints. Mobicent \cite{chen2010mobicent} is a credit-based incentive system which integrates credit and cryptographic technique to address  edge insertion and edge
hiding attacks among nodes. PI \cite{lu2010pi} proposes the inclusion of an
incentive on the  bundle sent by the source to incentivize   selfish nodes to cooperate in message delivery. 

Game theory is one of the common tools used to study 
strategic behavior.  
Ning et al. \cite{ning2011incentive} proposed a credit-based incentive scheme to promote nodal collaboration in a DTN with multiple interest types; they assume that a message may be desired by multiple destinations. The authors formulate nodal communication as a two-person cooperative game. 
  Wei et al. \cite{wei2011mobigame} proposed a user-centric  reputation based incentive protocol. They defined a game-theoretic framework to design  costs that leads to a Perfect Bayesian Equilibrium.   

Reputation mechanisms involving social networks have also been considered to cope with selfish behavior.  MobiGame \cite{wei2011mobigame} is a user-centric and social-aware reputation based incentive scheme for DTNs.  \cite{li2010routing} proposes socially selfish routing in DTNs, where a node exploits social willingness to determine whether or not to relay packets for others. 
In \cite{wang2012incentive}, authors proposed an incentive driven dissemination scheme that encourages
nodes to cooperate and chooses delivery paths that can reach as
many nodes as possible with fewest transmissions.

 A fundamental aspect that is usually ignored in the DTN literature concerns the challenge to transmit feedback messages.  In DTNs, feedback messages  may experience  large delays   
and for this reason  the exchange of rewards between relays should not require feedback to other nodes. To overcome such a challenge, the mechanism proposed in \cite{TMC16}, and also considered  in this paper,  assumes that  a relay receives a positive reward if and only if it is the first  to deliver the message to the  destination.   Similar ideas have been considered in \cite{chahin2013incentive}, where  a credit-based incentive system using the theory of minority games is adopted  to attain coordination in a distributed fashion.   The mechanism considers costs for taking part in the forwarding process which vary with  device technologies or users habits.

The  mechanism  investigated in this paper is a generalization of the one studied in \cite{ TMC16}  in which the relay nodes were assumed to be homogeneous with identical statistical model of mobility. 
Motivated by the fact that a realistic mobility is heterogeneous~\cite{romaTaxi}, we generalize the results in \cite{ TMC16} to heterogenous DTN network where different pairs of nodes might meet at different rates.

\subsection{Summary of the Contributions}


\textbf{Incentive mechanisms under heterogenous settings} (Section \ref{sec:ExpectedReward}):  we propose a model to enable incentives in DTNs accounting for relays that can have different inter-contact time statistics, under multiple information sharing structures. We show that the average reward paid by the source is independent of the  information sharing setting and depends upon the mobility pattern only through the mean inter-contact times. 

For exponentially distributed inter-contact times, we also give expressions for the reward the source must promise to a relay, as a function of the instant of time at which the encounter occurs  and the information sharing structure under consideration.

\textbf{TTL analysis} (Section \ref{sec:TTL}): we show  that nodes equipped with TTL caches can efficiently tune the TTL so as to trade between the costs of actively searching for the destination and the probability to  satisfy the source demands.

\textbf{Trace driven validation} (Section \ref{sec:simulation}): we validate and parameterize the proposed model using real traces  collected from taxi cabs in the city of Rome  \cite{romaTaxi}. One of the main observations we make from these trace-driven experiments is that our results are robust against variations in the statistical distribution of the mobility pattern. Specifically, even though the actual inter-contact distribution may not be known, one can compute the rewards assuming an exponential distribution, and this will still result in a fair outcome for the relays and for the source. By fair, we mean that neither the relay nor the source gain  by making the  assumption of exponential inter-contact times. This experimental evidence suggests that it is sufficient to know the mean inter-contact times for this mechanism to work.

 The rest of this paper is organized as follows: in the next section we introduce the system model and our working assumptions. In Section \ref{sec:ExpectedReward} we present the main result - the average reward paid by the source for a message is independent of the information sharing setting and depends upon the mobility pattern only through the mean inter-contact times.  In Section \ref{sec:TTL}, we characterize the effect of the time to live (TTL) counter on the delivery probability and energy storage. Section  \ref{sec:simulation} provides  simulation results, performed using both synthetic as well as real traces, that validate our analytical findings. Section \ref{sec:discuss} discusses the assumptions and the potential improvements to the model and Section~\ref{sec:conclusion} concludes. 

\section{SYSTEM MODEL}
\label{sec:sytemModel}
Consider a heterogenous DTN wherein 
nodes are equipped with    wireless  interfaces  allowing  communication  with  other  mobiles  in  their  proximity.  
The network comprises a source node, a destination node and $N$ mobile relays.    The set of relay nodes is denoted by   $\mathcal{R}=\{ r_1, \ldots, r_N \}$, where  $r_k$ refers to the  node with identifier $k$.  Assume that the source and the destination are fixed and that they are not in  range of each other. The source  seeks help from relays  to send its messages.   
A {\it contact} occurs   when two nodes move  within  the  transmission  range  of each  other.  
 As   the  density of  relay nodes  is assumed to   be  sparse,   we  consider    contacts exclusively between relays and the source or  the destination. Such routing strategy is referred to as {\it two-hop routing},  as messages  that reach the destination traverse two-hops.  

 To incentivize relay transmissions, the source promises to each relay it meets a (time-varying) reward with the condition that only the first relay to deliver the message to the destination will get its reward.   
When a contact between a relay and the source occurs, the relay can choose to participate in delivering the message. If the relay chooses  to participate, it  receives the message and incurs an energy reception cost denoted by $C_r$. The reception cost $C_r$ is fixed and  is the same for all relays. The relay should have enough buffer space to store the message and carry it until encountering the destination. Let $C_s$ be the energy cost  per time unit 
 for storing    and carrying the message. Finally, if a relay is the first to meet the destination among all relays that have the message, it delivers the message and receives the reward promised by the source. Sending the message to the destination incurs a transmission cost denoted by $C_d$. Hence, the reward proposed by the source has to, at least, offset the expected cost as estimated by the relay to deliver the message to the destination.

 The source can use a number of different mechanisms to implement micropayment to relays.  For instance, it can issue  electronic cheques encrypted with the public-key of the destination. The first relay who delivers the message sends the e-cheque to the destination who decrypts it  and returns it to the relay.


 


\subsection{The Role of Information}

Relay nodes estimate  costs  and revenues associated to the delivery of messages to the destination based on  the expected time  to reach the destination and the probability of success (i.e. the probability to be the first one to deliver the message to the destination). The success probability, in turn,  depends on 1) how many relays have already  joined the transmission efforts ({\it counts}), 2) their  {\it identities}  and 3) the {\it time instants} at which they have met the source.

Aiming towards reducing its costs, the source can act strategically,  and choose  the amount of  information to share with   relays.  
 We consider four information sharing strategies, with increasing levels of information provided by the source to the relays.  The strategies are summarized in Table~\ref{tab:infosharing}, and vary according to the three factors that determine the success probability, as discussed in the previous paragraph.           
In our analysis  we assume  that the  information sharing strategy  is fixed and given and does not discriminate between relays or between the times at which the source meets the relays. 
In addition, in all cases we  assume that  relays know  the inter-contact time  distributions between each relay,  the source and the destination, i.e., statistical information is common knowledge. In section \ref{sec:simulation}, it will be shown using simulations that knowledge of the inter-contact times is not essential. It is sufficient to know the mean inter-contact times in order to compute the rewards in a fair manner.

\begin{table}
\caption{Available information under different  sharing strategies} 
\begin{tabular}{|l||l|l|l|l|} 
\hline
Information  & statistics  & \multicolumn{3}{c|}{previous contacts info.} \\
sharing strategy   &   (CDFs) & counts & identities   & instants   \\
\hline
\hline
Full  $(F)$ & \checkmark & \checkmark &\checkmark &\checkmark  \\ 
\hline
Partial with  &         &            &          &    \\
 identities $(P+)$ & \checkmark & \checkmark &\checkmark & \\
\hline
Partial without  &         &            &          &    \\
 identities $(P-)$ &  \checkmark & \checkmark & & \\
\hline
None $(N)$ & \checkmark & & &  \\
\hline
\end{tabular} \label{tab:infosharing}
\end{table}




To appreciate some of the implications of the different information settings, we compare the payments  requested by the relays to the source under the different  settings.     
Under the full information setting, if a relay knows that there are many copies of a message in circulation, it will  infer that it has a higher risk of failure compared to  if it were the first one to meet the source. Therefore, the larger the number of copies of a message in circulation, the  higher the reward  requested by the relay.   
Under the no information setting, in contrast,  the first relay who meets the source will certainly underestimate its probability of success and ask for a higher reward compared to the setup wherein it knew it  were the first  to encounter the source.

A natural question which we will address in Section~\ref{sec:ExpectedReward} refers to  the optimal information sharing strategy from the perspective of a strategic  source who aims towards minimizing   its costs.  As suggested in the paragraph above, the full information setting favors higher payments to late arrivals, whereas the no information setting favors newcomers.  Interestingly, we will show that the balance between newcomers and late arrivals implies that {\it the  expected payment incurred by the source  is the same in all the considered settings}.

\label{sec:balance}

\subsection{Mobility Patterns}
Next, we introduce the considered mobility  model.  
Let $\tilde{T}^i_s$ (resp. $\tilde{T}^i_d$) be a random variable characterizing the  time between any two consecutive contacts between  relay $i$  and the source (resp. the destination). We assume  that the inter-contact times between  relays and the source (resp. the destination) are independent and non-identically distributed. In addition, we assume  that the contacts between relays and the source or the destination are instantaneous, i.e., the duration of these contacts can be neglected.

The instant at which a message is generated by the source can be seen as a random point in time with respect to the contact process of a relay with the source. Hence, the random time between the instant at which the message is generated and the instant at which  relay $i$ will meet the source corresponds to the residual life of its inter-contact time, which we denote by $T^i_s$. Similarly, 
the time instant at which  relay $i$ meets the destination  after contacting the source is a random point in time with respect to the contact process of this relay with the destination.  
Hence,  the time to meet the destination is given by the residual life of the inter-contact
time distribution with the destination, and is denoted by $T^i_d$.
The complementary cumulative distribution function (CCDF) of $T_s^i$ (resp. $T_d^i$) is denoted by $F_s^i(x) = \mathbb{P}(T_s^i>x)$ (resp. $F_d^i(x) = \mathbb{P}(T_d^i>x)$). 
The mean residual inter-contact time of relay $i$ with the source and the destination is given by 
$\mathbb{E}[{T}_s^i] = {\mathbb{E}[({{\tilde{T}}_s^i})^2]}/({2\mathbb{E}[{\tilde{T}}_s^i]})$ and  $\mathbb{E}[{T}_d^i] = {\mathbb{E}[({{\tilde{T}}_d^i})^2]}/({2\mathbb{E}[{\tilde{T}}_d^i]})$, respectively.

\section{Expected Costs and Rewards} 
\label{sec:ExpectedReward}
In this section, we shall first compute the probability of success as estimated by a  relay that meets the source.  
Then, we  compute the reward that this relay will ask the source, so as to assist in the transmission of the message. Finally, we shall investigate the cost incurred by the source for 
transmitting a message to the destination. Both these quantities will be studied in all the four information settings.  All proofs are available in our technical report \cite{report}.

\subsection{Probability of Success}
Let $s_i$, $i=1,..,N$, be the random time at which the source meets the $i$-th relay.  We denote by ${\bss}$ the vector $(s_1,..,s_N)$.  In order to simply the notation, we shall write  ${\bss_{-n}}$ to denote the vector $(s_1,..,s_{n-1},s_{n+1},..,s_N)$ and ${\bss_{m:n}}=(s_m,..,s_n)$. 

Let $L$ be an ordered set  comprising  all the possible orderings of the $N$ nodes in the system, $|L|=N!$.
  The elements in $L$ are ordered in lexicographic order.
Let $\bm{\ell} \in L$ denote an ordering of the $N$ nodes in the system, and let $\bm{\ell}_m \in L$ be the
 $m$-th ordering.   
Then, $\ell(i)=r_k$ if the $i$-th    node to meet the source under ordering $\bmell$ is node $r_k$. 
  Node $\ell(i)$ meets the source at time $s_i$. We denote by $\ell^{-1}(r_k)$ the index of relay $r_k$ in vector 
  $\ell$, i.e.,  $\ell^{-1}(r_k)=i$ if  $\ell(i)=r_k$ under  ordering $\bmell$.   To simplify notation, we shall write  ${\bf \bmell_{-n}}$ to denote the vector $(\bmell_1,..,\bmell_{n-1},\bmell_{n+1},..,\bmell_N)$ and ${\bf \bmell_{m:n}}=(\bmell_m,..,\bmell_n)$. 


   
  Let $p_{j}^{(k)}(\bss, \bmell)$ be the success probability estimated by the $j$-th relay to meet the source
   under setting $k\in\{F, P+, P-, N\}$.    Let
   $p_{j}(\bss, \bmell)$ be the actual success probability of relay $\ell(j)$ accounting for all contact times. 
   The probability of success of $r_k$ under the full information setting given $\bss$ and $\ell$
 is given by $ p_{\ell^{-1}(r_k)}^{(F)}(\bss, \bmell)$.  Recall
  that $p_{j}^{(F)}(\bss, \bmell)$  depends only  on ${\bss}_{1:j}$ and ${\bf \bmell}_{1:j}$.  Similar notation is used for the other settings.  
  To simplify presentation, with some abuse of notation we may drop  elements from $\bss$ whenever they are unnecessary. 
The first and last nodes to meet the source satisfy the following relations,
\begin{eqnarray}
&&p_{1}^{(P+)}(s_1, \bmell) =p_{1}^{(P-)}(s_1, \bmell) =p_{1}^{(F)}(s_1, \bmell) \label{meet1}, \\
&&\hspace{0.35cm}p_{N}^{(F)}(\bss, \bmell) = p_{N}(\bss, \bmell).  \label{meet2}
\end{eqnarray}
The first equality in  \eqref{meet1}  follows from the fact that  the first relay to meet the source obtains  the same
information   under the partial  and
 full information settings. Similarly,~\eqref{meet2} follows from the fact that in the full information setting, the last relay knows the
contact times of all other relays with the source.

Let $f^{\source}_{r_k}(x)$ be the pdf of the meeting time between $r_k$ and the source, and let $f^{\source}_{i}{(x,\bmell)}$ be the pdf of the meeting time between the  $i$-th relay in $\bmell$ and the source. Given $\bmell$, we have 
$f^{\source}_{i}{(x,\bmell)}=f^{\source}_{r_{\ell(i)}}{(x)}$.  
Then,
\begin{eqnarray}
f^{\source}(\bss,\bmell)= \prod_{i=1}^N f^{\source}_{i}(s_i,\bmell), \;\; f^{\source}(\bss_{j:k},\bmell)= \prod_{i=j}^k f^{\source}_{i}(s_i,\bmell)\\
f(\bss_{1:N}) =\sum_{m=1}^{N!} f(\bss_{1:N}, \bmell_m)\hspace{03.2cm}
\end{eqnarray}
Let $f^{\source}_{i}{(\bss,\bmell)}$ be the pdf of the meeting time between the source and the 
  $i$-th relay in $\bmell$, 
conditioned that the first $i-1$ encounters between the source and  relays have already occurred.  We let
\begin{eqnarray}
f^{\source}_{i:j}{(\bss,\bmell)} = \frac{\prod_{k=i}^j f^{\source}_{k}(s_k,\bmell)}{\int_{s_{i+1}=s_i:s_j=s_{j-1}}^\infty \prod_{k=i}^j f^{\source}_{k}(s_k,\bmell)ds_{j:i+1}}
\end{eqnarray}
Then, if $j=i$ we have
\begin{eqnarray}
f^{\source}_{i}{(\bss,\bmell)} = \frac{ f^{\source}_{i}(s_i,\bmell)}{1-F_i(s_{i-1},\bmell)} 
\end{eqnarray}


If inter-contact times are exponentially distributed, the following proposition provides closed-form  expressions for the probability of success.  Let $\lambda_i$ (resp., $\mu_i$) be the inter-contact rate  between relay $i$ and the source (resp., the destination).   
\begin{proposition}[See the proofs in appendix A] 
\label{propo:exponential}
\textbf{Full information setting:} given information about previous inter-contact time instants through the first $n$ entries of  $\textbf{s}$ and ${\bmell}$,  
we have
\begin{multline}
p_{\ell(n)}^{(F)}(s_n) = \\
 \prod_{k=1}^{n-1} e^{-\mu_{\ell(k)}(s_n-s_k)} 
 \sum_{i=n}^{N} \frac{\mu_{\ell(n)}}{\sum_{l=i+1}^N \lambda_{\ell(l)}  + \sum_{l=1}^i \mu_{\ell(l)}} \times \\ 
 \prod_{k=n+1}^{i} \sum_{j=k}^{N} \frac{\lambda_{\ell(j)}}{ \sum_{l=k}^N \lambda_{\ell(l)} + \sum_{l=1}^{k-1} \mu_{\ell(l)})}  \nonumber
\end{multline}
\textbf{Partial information setting with identity information:} given the number and the identity of relays that previously contacted the source, we have
\begin{multline}
p_n^{(P^+)}(s_n) = \prod_{i=1}^{n-1}\lambda_{i}\psi(i,n) \times \\
\sum_{i=n}^{N}  \prod_{k=n+1}^{i} \sum_{j=k}^{N}\frac{\lambda_j}{\sum_{l=k}^N \lambda_l + \sum_{l=1}^{k-1} \mu_l}  \frac{\mu_n}{\sum_{l=i+1}^N \lambda_{l} + \sum_{l=1}^i \mu_l } \nonumber
\end{multline}
where
\begin{equation}
\psi(i,n)=\left\{
\begin{array}{ll}
\frac{(e^{-\mu_{i}s_n} - e^{-\lambda_{i}s_n})}{(\lambda_{i}-\mu_{i})(1-e^{-\lambda_{i}s_n})}, & \textrm{if } \lambda_i \neq \mu_i, \\
 s_n {e^{-\lambda_i s_n}}/({1-e^{-\lambda_i s_n}}), & \textrm{otherwise} \nonumber
 \end{array}
 \right.
 \end{equation}
\textbf{Partial information setting without identity information:} given only the number of relays that previously contacted the source, $n-1$, we have
\begin{multline} 
p_n^{(P^-)}(s_n) = \sum_{i=1}^{C_{N-1}^{n-1}} \prod_{j=1}^{n-1}\lambda_{\ell_i(j)}\psi(\ell_i(j),n) \prod_{k=n+1}^{N} e^{-\lambda_{\ell_i(k)}s_n} \times \\
\sum_{i=n}^{N}  \prod_{k=n+1}^{i} \sum_{j=k}^{N}\frac{\lambda_j}{\sum_{l=k}^N \lambda_l + \sum_{l=1}^{k-1} \mu_l}  \frac{\mu_n}{\sum_{l=i+1}^N \lambda_{l} + \sum_{l=1}^i \mu_l } \nonumber
\end{multline}
where
\begin{equation}
\psi(\ell_i(j),n)=\left\{
\begin{array}{ll}
\frac{(e^{-\mu_{\ell_i(j)}s_n} - e^{-\lambda_{\ell_i(j)}s_n})}{(\lambda_{\ell_i(j)}-\mu_{\ell_i(j)})(1-e^{-\lambda_{\ell_i(j)}s_n})}, & \hspace*{-7mm} \textrm{if } \lambda_{\ell_i(j)} \neq \mu_{\ell_i(j)}, \\
 s_n {e^{-\lambda_{\ell_i(j)} s_n}}/({1-e^{-\lambda_{\ell_i(j)} s_n}}), & \textrm{otherwise} \nonumber
 \end{array}
 \right.
 \end{equation}
\textbf{No information setting:} given solely statistical information about inter-contact times, we have  
\begin{equation}
p_n^{(N)}(s) = \sum_{m=1}^{N} \sum_{i=1}^{C_{N-1}^{m-1}} \prod_{j=1}^{m-1} (1-e^{-\lambda_{\ell_i(j)}}) \prod_{k=m+1}^{N} e^{-\lambda_{\ell_i(k)}} p_m^{(P)}(s) \nonumber
\end{equation}
\end{proposition}

\subsection{Expected Cost for a Relay}
Next, our goal is to compute the expected cost estimated by a relay to transmit a packet. Let $V_i^{(k)}(\bss,\bmell)$  be  the net expected cost incurred  to transmit a packet to the destination, as estimated by the  $i$-th node in $\bmell$
  under setting $k$. Let $R_i^{(k)}(\bss,\bmell)$
be the expected reward offered by the source to the  $i$-th node in $\bmell$ to transmit a packet to the destination under setting $k$, 
$k \in \{ F, P+,P-, N\}$.  Recall that the reward is granted only to successful nodes. Then,
\begin{equation} \label{eq:netcost}
V_{i}^{(k)}(\bss, \bmell) = c_r + c_s \mathbb{E}(T_d^{\ell(i)}) + (c_d - R_i^{(k)}(\bss,\bmell)) p_i^{(k)}(\bss, \bmell)
\end{equation}

The first term in the net expected cost~\eqref{eq:netcost} is the reception cost, which is always incurred. The second term represents the expected in transit (storage) cost. It is directly proportional to the mean of the residual inter-contact time between relay $\ell(i)$ and the destination. The last term is the cost of transmitting the message to the destination which in turn confers a reward to the relay. This term enters into play only if relay $\ell(i)$  is the first one to reach the destination, which explains the factor $p_i^{(k)}(\bss, \bmell)$.


\subsection{Expected Reward Paid by the Source}

Relay $r_k$ will accept to forward a message to the destination  if  the reward promised by the source offsets it expected cost, that is, if $R_i^{(k)}(\bss,\bmell) $
is such that  $\mathbb{E}(V_{i}^{(k)}(\bss, \bmell) )\leq 0$. Thus the minimum reward that the source has to promise relay $r_k=\bmell^{(-1)}(i)$  is 
\begin{eqnarray}
 R_i^{(k)}(\bss,\bmell) &=& c_d + \mathcal{C}_{\ell{(i)}} / p_i^{(k)}(\bss, \bmell)
\end{eqnarray}
where  $\mathcal{C}_{\ell{(i)}}=c_r + c_s E(T_d^{\ell{(i)}})$. Thus we have 
\begin{equation}R^{(k)}(\bss,\bmell) = c_d + \sum_{i=1}^{N} \mathcal{C}_{\ell{(i)}}  p_i(\bss, \bmell) / p_i^{(k)}(\bss, \bmell).
\end{equation}
Note that the reward requested by relay  $r_k$ depends on the information given by the source only through the estimated probability of success $p^{(k)}_{j}(\bss, \bmell)$. 

Now we turn our attention to the expected reward paid  by the source when the expectation is taken over all possible meeting times. This quantity can be thought of as the long-run average reward per message the source will have to pay if it sends a large number of messages (and assuming that message generation occurs at a much slower time scale than that of the contact process). 

\begin{lemma}[See the proofs in appendix B]
\label{lemma:ExpectedRewardPaidByTheSource}
The expected reward paid by the source  under setting $(k)$ is  given by 
\begin{align}
& R^{(k)} =   \nonumber  \\
&  c_d + \sum_{\bmell_m \in L} \sum_{i=1}^{N} \mathcal{C}_{\bmell_m{(i)}}  \int_{s_{1}=0:s_i=s_{i-1}}^{\infty} \frac{  f(\bss_{1:i}, \bmell_m)}{ p_i^{(k)}(\bss, \bmell_m)} \varphi_i(\bss,\bmell_m) ds_{{i:1}} \nonumber \\ \label{eq:averageRewardGenralCase}
\end{align}
where
\begin{equation}
\varphi_i(\bss,\bmell_m) =  \int_{s_{i+1}=s_i:s_N=s_{N-1}}^{\infty} \hspace{-10mm} p_i(\bss, \bmell_m) f(\bss_{i+1:N},\bmell_m) ds_{{N:i+1}} 
\end{equation}
\end{lemma}


From the probability of success estimated by the relays in the four settings, we can prove that the expected reward paid by the source for delivering its message is the same in all four settings, as stated in Theorem \ref{theorem1}.
\begin{theorem}[See the proofs in appendix C]
\label{theorem1}
The expected reward to be paid by the source under setting  $k\in \{F, P^{+}, P^{-}, N\}$ is
\begin{equation}
R^{(k)}= c_d + \sum_{i=1}^{N} \mathcal{C}_{r_i}
\end{equation}
\end{theorem}

Theorem \ref{theorem1} shows that  the excepted reward paid by the source  remains the same irrespective of the information it conveys.  As mentioned in Section~\ref{sec:balance}, this is due to  the  balance of the flow between late arrivals and newcomers: the former are favored by  the  sharing of information, whereas the later 
benefit from information concealing.

\section{The effect of  Time To Live (TTL)} \label{sec:TTL}
Next, we consider the scenario wherein  relays stop trying to forward a packet 
after a prescribed period of time.   To this aim, we associate to each node and to each 
packet a time to live (TTL) counter.  After this time is elapsed, the message will be dropped.  The counter associated to  node $i$ ticks at rate $\mu'_i$. We further  assume
that the time between ticks is exponentially distributed.  By the time the timer at node $i$ ticks, 
its packet is dropped
and the node enters into sleep mode.  

In what follow, we study the impact of the TTL counter accounting for
 the tradeoff between energy gains and the probability that the destination does not receive the packet of interest. We  address the following question:  \emph{what is the impact of the TTL on the delivery probability and on the energy spent by relays in order to transmit  the message?}

We consider  a single source-destination pair,  and assume exponentially distributed inter-contact times.
 Future work consists of extending the analysis to  relays equipped with  TTL caches that transmit multiple packets between different source-destination pairs.

Recall that $\mu_i$ is the rate at which relay $i$ meets the destination.   Let $\mu'_i = \mu_i \epsilon$, with $0 < \epsilon < 1$.
 Let $W_i$ be the time it takes for relay $i$ to either meet the destination or drop a packet when its TTL expires, whatever
 occurs first. Then, $W_i$ is an exponentially distributed random variable, with rate $\mu_i ( 1+\epsilon)$.
Let $D$ be the probability that the destination does not get the packet of interest, assuming that the source sets incentives so that all relays opt to assist in the transmission, 
\begin{equation}
D = \prod_{i=1}^N \frac{\epsilon \mu_i}{\epsilon \mu_i + \mu_i} = \left(\frac{\epsilon }{\epsilon  + 1}\right)^N 
\end{equation}
Then,
$\epsilon = {D^{1/N}}/({1-D^{1/N}})$.

 Let $\rho_i$ be the mean reduction in the length of  time   during which relay $i$ tries to transmit a packet, due to  the TTL counter, 
\begin{equation}
\rho_i=E[T_d^i]-E[\tilde T_d^i]= \frac{1}{\mu_i} - \frac{1}{\mu_i ( 1+\epsilon)}  = \frac{\epsilon }{1+\epsilon} E[T_d^i]
\end{equation}
where $T_d^i$ and $\tilde T_d^i$ are the times invested in transmitting a packet before and after equipping  node $i$ with TTL counters, respectively.  Let $G$ be the relative gain  of a relay node $i$ in terms of the expected in transit (storage) cost due to the introduction of the  TTL counter. Thus
\begin{equation}
G=  \frac{\ c_sE[T_d^i]-c_s E[\tilde T_d^i]}{c_s E[T_d^i]} = \frac{\epsilon }{1+\epsilon} 
\end{equation}
Note that neither $D$ nor $G$ depend on 
the rates at which relays meet the destination. 
%
We write $D(N,G)$ as a function of $N$ and $G$,
%
$D(N,G) = G^N ,   0 \leq G < 1$.  Therefore, a linear increase  in  $G$  yields a sub-linear increase in 
 $D(N,G)$.  


\section{Simulation Results}
\label{sec:simulation}
To validate our theoretical results, we run simulations using two types of mobility traces: $(i)$ synthetic traces generated using the ONE (Opportunistic Network Environment) simulator \cite{ONE} and $(ii)$ real traces collected from taxi cabs in the city of Rome  \cite{romaTaxi}.
Our goals are to $(a)$ parameterize the inter-contact times using real data and validate our analytical model, $(b)$ assess the average reward set by the source under a realistic setting and $(c)$ estimate the instantaneous rewards received by relays over time.

The key metrics that we extract from our traces are  the inter-contact time distributions  between relays and the source, and between relays and  the destination.  
In our simulations, when the source meets a relay it uses the inter-contact time distribution between relays and the destination to compute the  success probability as estimated by the relay.  This  probability of success is then used to obtain the  reward that  the source promises to the relay.  In addition, from the traces we determine which  relay  gets the reward from the destination,  for each message generated by the source.    Finally, we make use of the mean   inter-contact time between relays and the destination to compute the desired metrics of interest,  
 using the theoretical results presented  in Section~\ref{sec:ExpectedReward}, and contrast them against our simulation results. 


\begin{figure}
\begin{center}
\includegraphics[width=0.9\columnwidth]{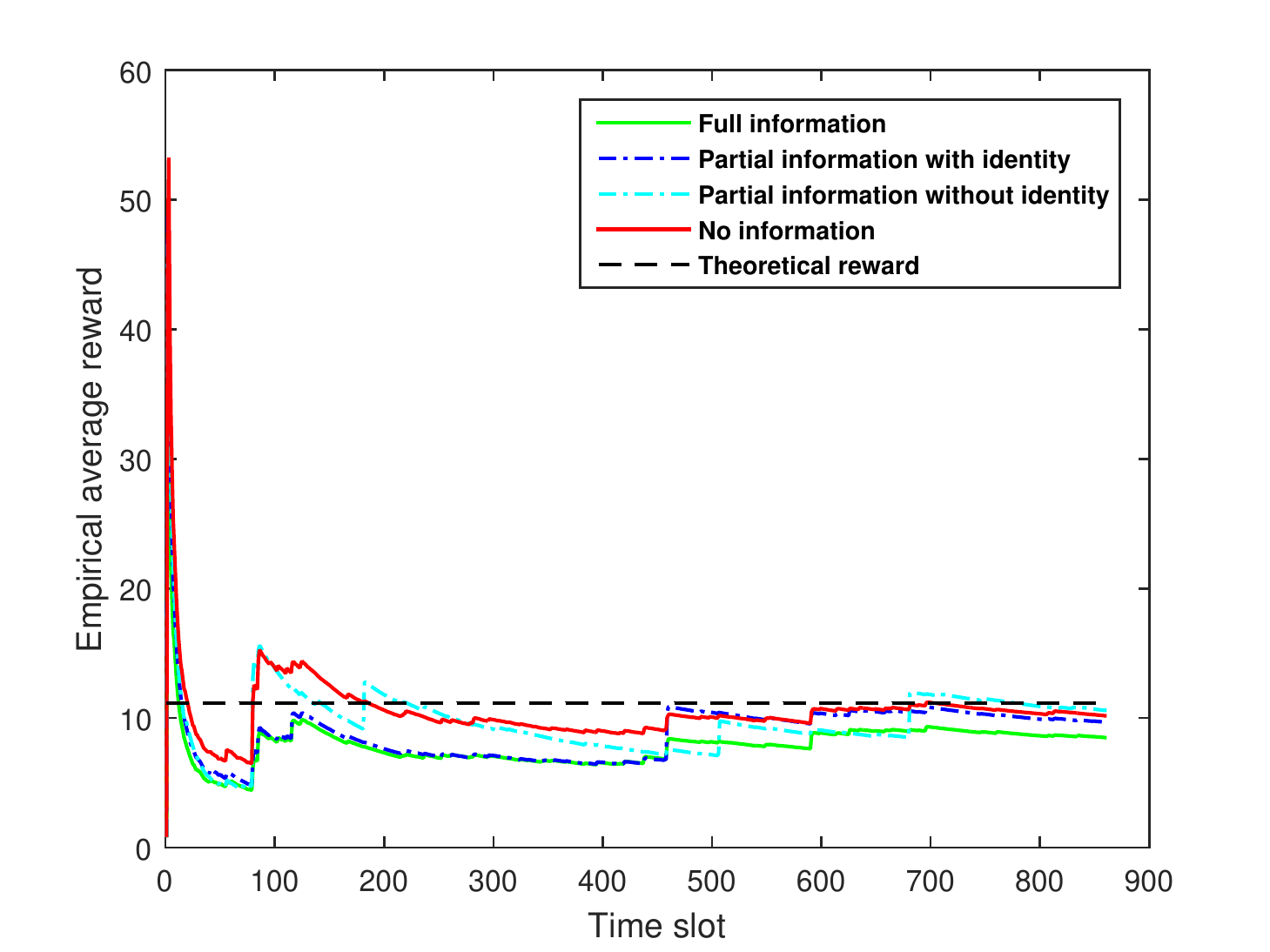}
\end{center}
\vspace{-0.2in}
\caption{Empirical average reward under the four information settings vs the theoretical reward With $N=10$, $C_d = 0.4$, $C_r=0.04$, $C_s = 0.01$} 
\label{fig:16_relays_rwp} \vspace{-0.2in}
\end{figure}

\begin{figure}
\begin{center}
\includegraphics[width=0.9\columnwidth]{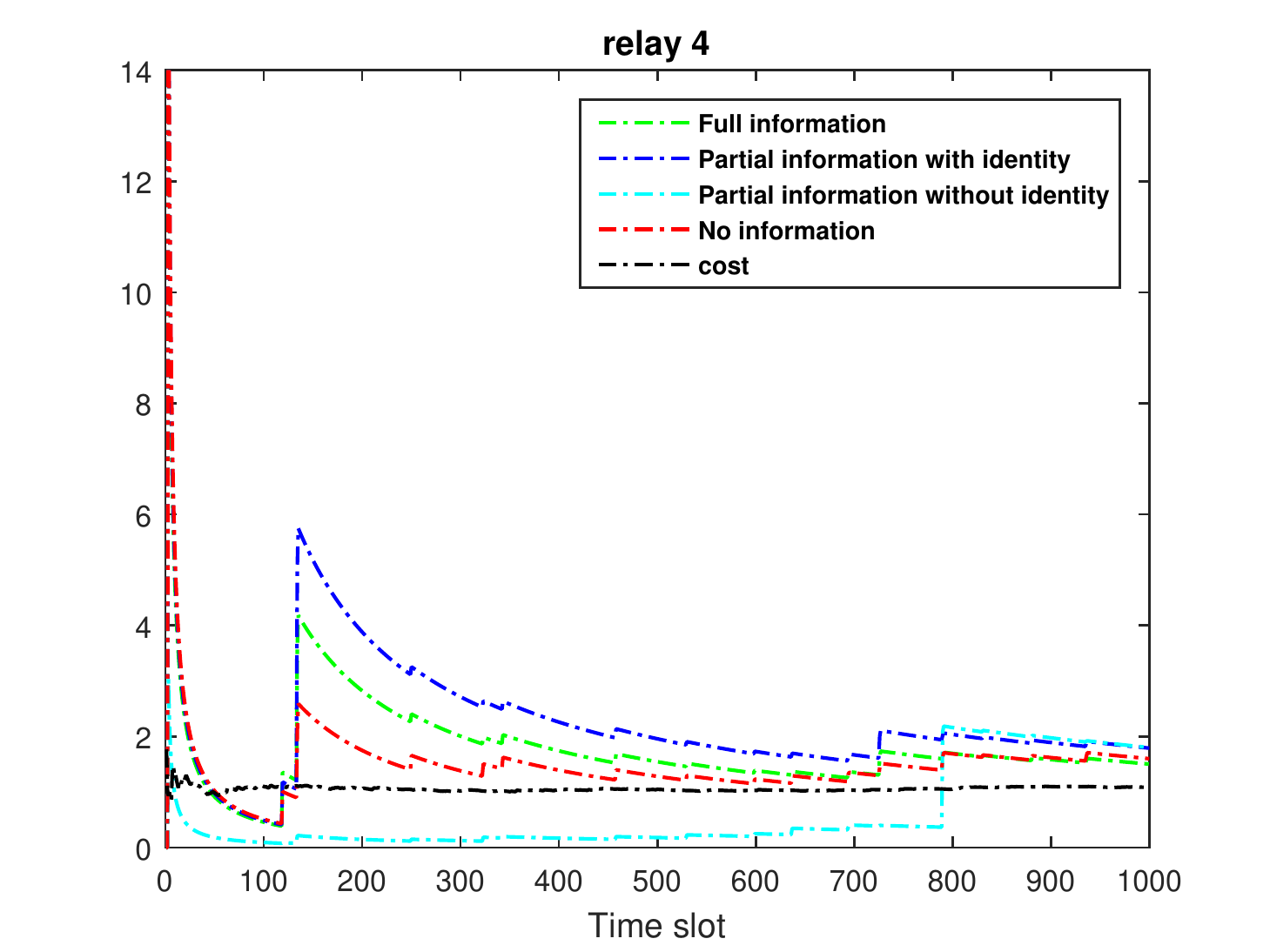}
\end{center}
\vspace{-0.2in}
\caption{Empirical average reward and the energy cost of a relay under the four information settings for $N=10$} 
\label{fig:instantanuous_reward_one_relay_rwp} \vspace{-0.2in}
\end{figure}

\subsection{Rewards and Costs under Synthetic Traces }
In our initial set of simulation results, we consider  mobility traces generated using  the Random Way Point mobility model (RWP).  In this setting, our simulations were conducted using the ONE simulator \cite{ONE}. Under the RWP model, each node moves independently to a randomly chosen destination with a randomly selected  velocity. Upon reaching the destination, the node waits for a period of time (known as pause time) before choosing another destination and velocity. We consider $N=10$ relays placed randomly in an area of $2500 \times 2500$~m$^2$. The transmission range of each relay is $10$ m. To take into account the heterogeneity, we have considered two classes of relays: the first class contains $10$ relays who choose their velocities and pause times uniformly at random in the ranges $[1, 13.9]$ m/s and $[0, 14400]$ s, respectively. The second class contains $6$ relays that choose their velocities and pause times  uniformly at random in the ranges $[0.5, 1.5]$ m/s and $[0, 3600]$~s, respectively. We collect  inter-contact times for a simulation time of over  $1$ year.  Once the inter-contact time distribution is generated, we  use it to compute the  reward promised to each relay at each contact  (refer to our  technical report~\cite{report} for further details on the inter-contact times and additional statistics obtained via  simulations). 


To compute the probability of success and the reward requested by a relay to deliver a message we use the results presented in Proposition~\ref{propo:exponential}  
  making the assumption that the inter-contact times between relays, the source and the destination are exponentially distributed.
%
Each message is generated by the source  every 10 hours. This
period of 10 hours will be called a time-slot. In Figure~\ref{fig:16_relays_rwp} we plot the analytical and the empirical average reward paid by the source as a function of time (measured in  time-slots) under the three information settings. The empirical average reward at  time-slot $t$ is  given by  
$\frac{1}{t} \sum_{n=1}^{t} R(n)$,
where  $R(i)$ is the reward paid by the source  for the message  generated at time-slot $n$.   
We observe that the empirical reward under the three information settings  converges to the theoretical reward given by Theorem~\ref{theorem1}. In addition, the convergence time of the full and partial information settings is faster when contrasted against the no information case.  

Next, we consider  the rewards and costs (of energy and storage) associated to the relays. For this purpose, we tag a relay chosen arbitrarily among the 10 relays, and plot the empirical average reward received by that relay from the source  as well as the associated  average  cost, as a function of the time-slot (see  Figure~\ref{fig:instantanuous_reward_one_relay_rwp}). We observe that the reward paid by the source in the four settings offsets the expected cost of that relay as predicted by the analytical model. 

\begin{figure}
\begin{center}
\includegraphics[width=0.9\columnwidth]{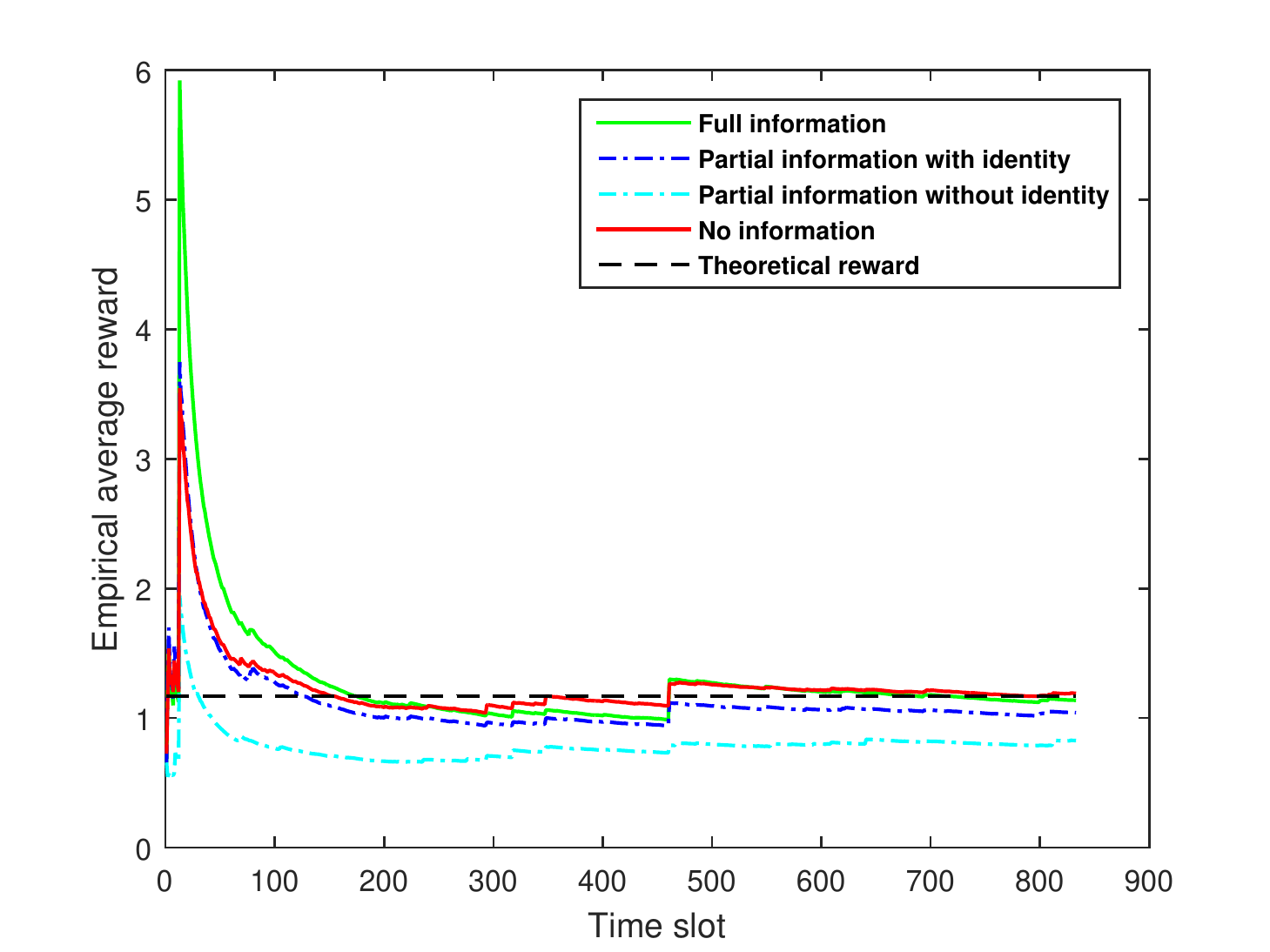}
\end{center}
\caption{Empirical average reward under the four information setting vs the theoretical reward With $N=10$, $C_d = 0.4$, $C_r=0.04$, $C_s = 0.01$} 
\label{fig:10_relays_mixt} \vspace{-0.2in}
\end{figure}

\subsection{Sensitivity Analysis and Robustness}
\label{ssec:robust}
\begin{figure}
\begin{center}
\includegraphics[width=0.9\columnwidth]{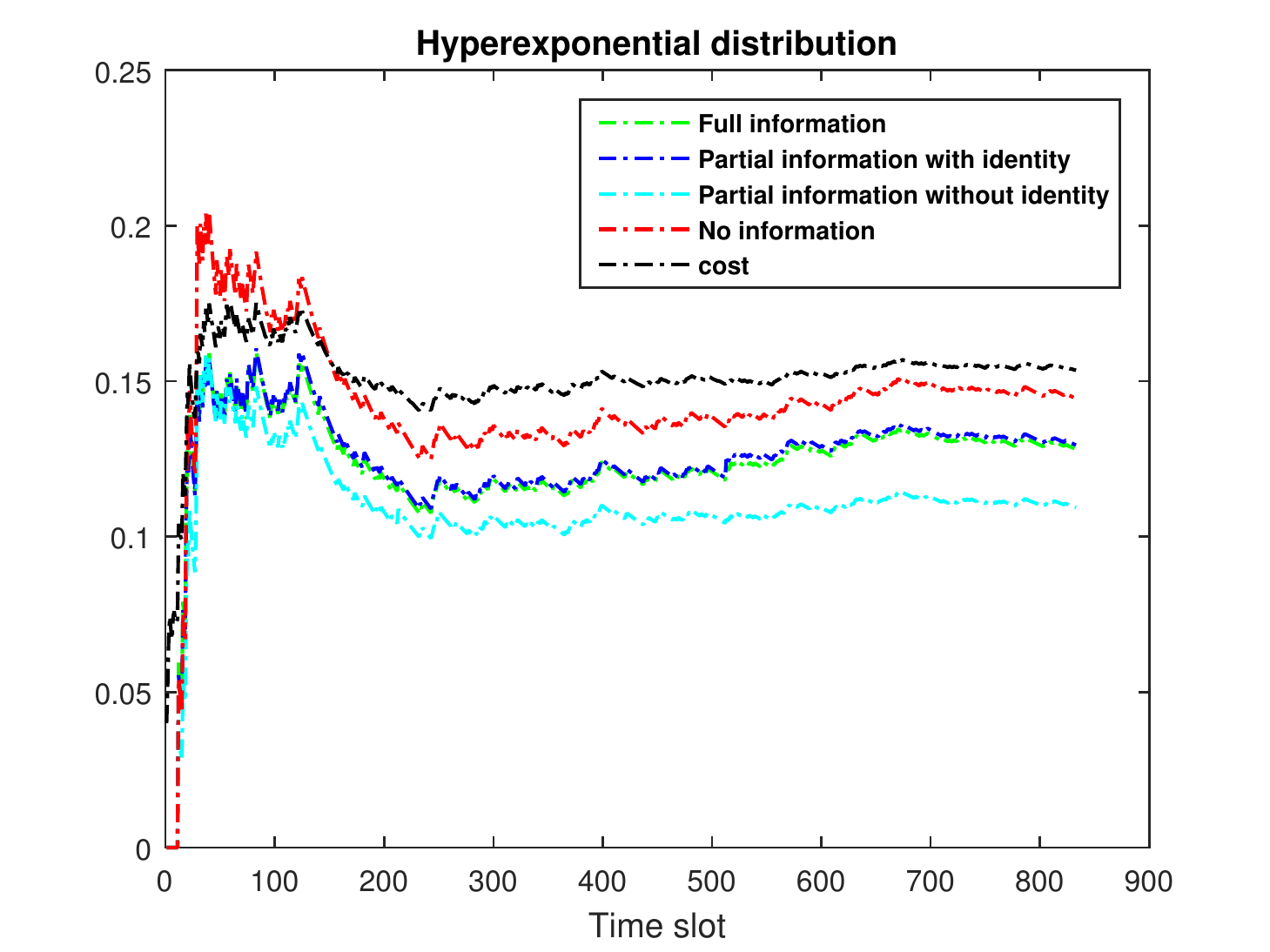}
\end{center}
\vspace{-0.2in}
\caption{Empirical average   of the reward and the energy cost for a relay with hyper exponential distribution of the inter-contact time   under four  information settings for $N=10$, $C_d = 0.4$, $C_r=0.04$, $C_s = 0.01$} 
\label{fig:instantanuous_reward_one_relay_hyper} \vspace{-0.2in}
\end{figure}

\begin{figure}
\begin{center}
\includegraphics[width=0.9\columnwidth]{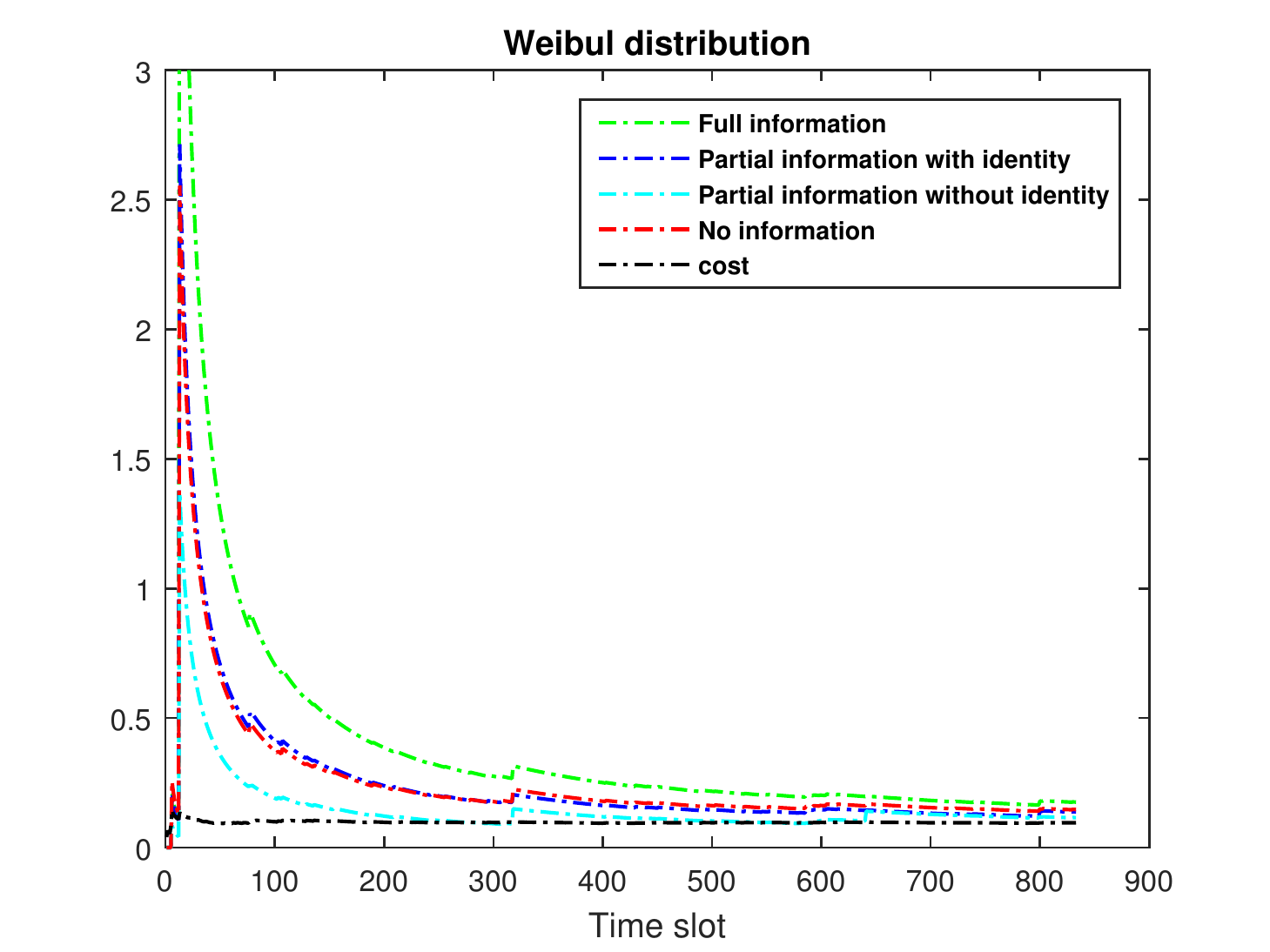}
\end{center}
\vspace{-0.2in}
\caption{Empirical average  reward and  energy cost for a relay with  the weibul  distribution  inter-contact time  under four  information settings for $N=10$, $C_d = 0.4$, $C_r=0.04$, $C_s = 0.01$} 
\label{fig:instantanuous_reward_one_relay_weibul} 
\end{figure}

Next, we further investigate the sensitivity of our results with respect to the assumption that  inter-contact times are exponentially distributed. To this aim, we 
 consider a source and relays that assess rewards using the exponential assumption even when the inter-contact times between relays and the source or destination follow other distributions.  As in the previous section,  we consider  $N=10$ relays, each of which associated to  a different inter-contact time distribution (see Table~\ref{exp-hyper-weib-nor}).  Figure~\ref{fig:10_relays_mixt} shows that the empirical reward under the four information settings  converges to the theoretical reward given by Theorem~\ref{theorem1}. 
Similarly,  in Figures~\ref{fig:instantanuous_reward_one_relay_hyper} and ~\ref{fig:instantanuous_reward_one_relay_weibul}  we consider 
%
hyper exponential and Weibull inter-contact distributions, respectively. 
 We note that the proposed reward mechanism remains 
   robust against changes in the mobility patterns.   In particular,  
   the observations made about  Figure~\ref{fig:10_relays_mixt} also hold for Figures~\ref{fig:instantanuous_reward_one_relay_hyper} and ~\ref{fig:instantanuous_reward_one_relay_weibul}:  
   the empirical rewards   converge to the theoretical values given by Theorem~\ref{theorem1}.
   

\begin{table}[h]
\vspace{-0.1in}
\caption{Contact rates of  relays with  source and  destination} \vspace{-0.1in}
\begin{center}
\begin{footnotesize}
\begin{tabular}{|c|c|c|c|}
\hline
 Relay & Distribution  & $\lambda$  &  $\mu$  \\
\hline
 r$_1$ &  Exponential &  0.6530  & 0.7945 \\
 \hline
 r$_2$ &  Exponential  &0.5296   &  0.2824 \\
 \hline
 r$_3$ & Hyperexponential &0.6714  &  0.6704 \\
 \hline
 r$_4$ &  Hyperexponential &0.6685 & 0.6670 \\
 \hline
 r$_5$ &  Weibull& 0.2483 &  0.2492  \\
  \hline
  r$_6$ &Weibull& 0.1647   &  0.1996\\
   \hline
   r$_{7}$ &Weibull&  0.2500    &  0.2000\\
    \hline
    r$_{8}$ &Folded normal &  0.1999  &  0.1991\\
     \hline
     r$_{9}$ &Folded normal&  0.2002   &  0.2015\\
      \hline
     r$_{10}$ &Folded normal &  0.1991  &   0.2005 \\
 \hline
\end{tabular}
\end{footnotesize}
\end{center}
\label{exp-hyper-weib-nor} \vspace{-0.1in}
\end{table}


\subsection{Real-World Traces}
\begin{figure}
\begin{center}
\includegraphics[scale=0.4]{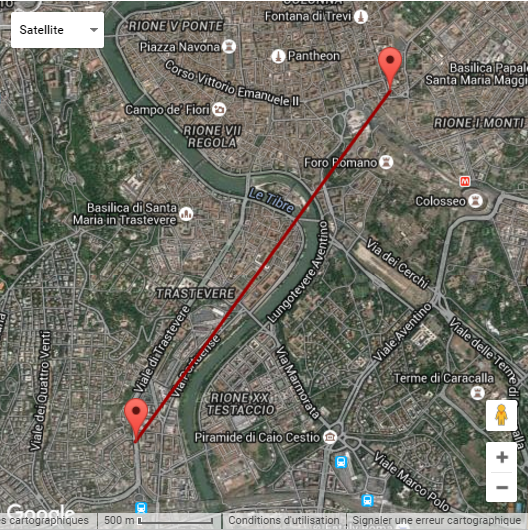}
\end{center}
\caption{Position of the source and the destination in the city of Rome} 
\label{fig:map_rome_satelitte}
\end{figure}

Next, we consider a dataset comprising  traces of taxis moving in the city of Rome. Each taxi is equipped with an Android tablet device running an application that updates the current position of the taxi every 7 seconds. This dataset contains GPS coordinates of approximately 320 taxis collected over 30 days. The traces were compiled in February 2014 \cite{romaTaxi}.
In this simulation scenario, we consider a source-destination
pair in the centre of Rome. 
The distance between the source and the destination is roughly $2.8$ Km. Using Google Maps, Figure~\ref{fig:map_rome_satelitte} shows the positions of the source and the destination. The transmission range of each taxi is $50$~m. 

\begin{figure}
\hspace{-0.3in}
\includegraphics[scale = 0.65]{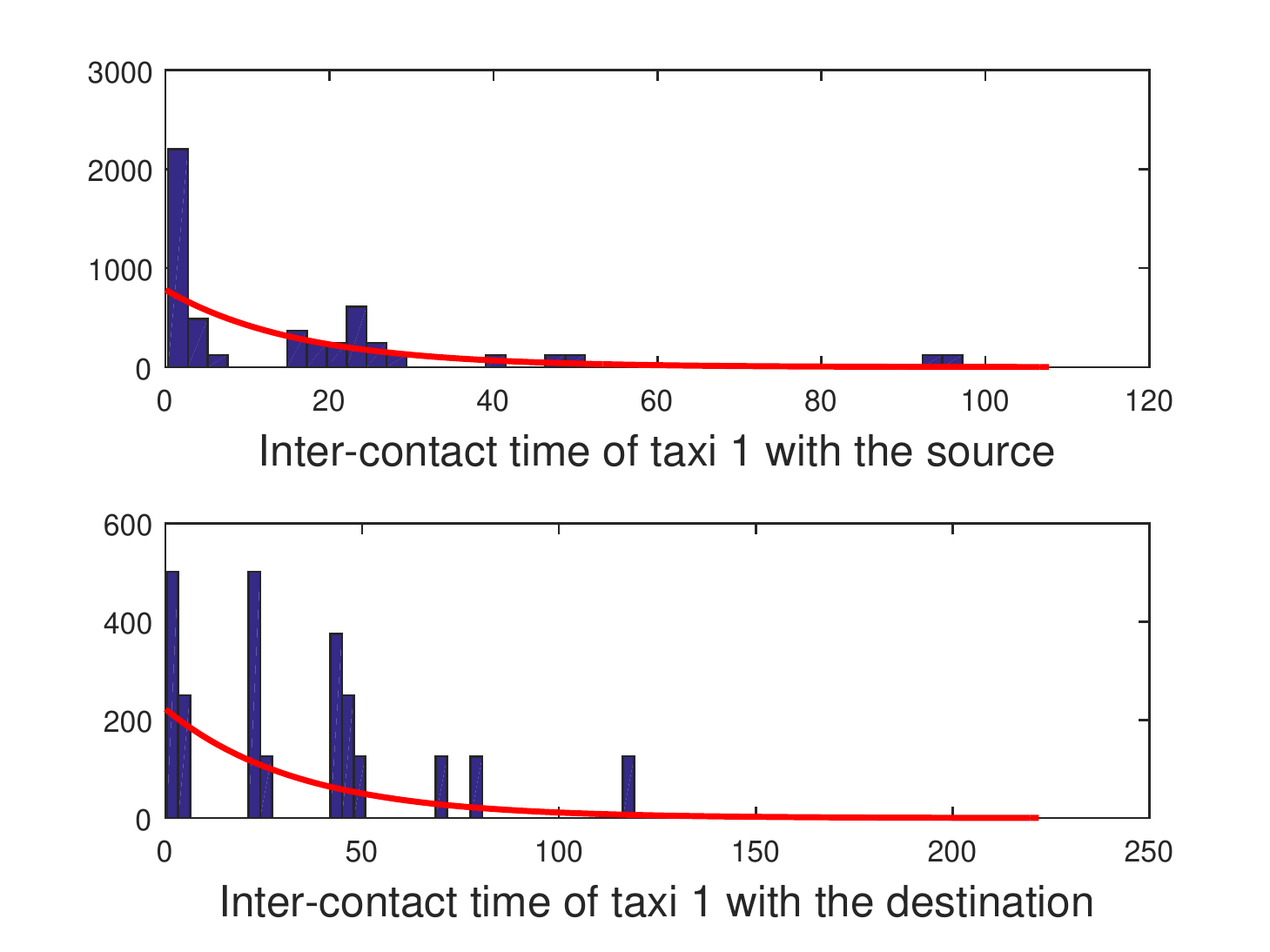}
\vspace{-0.2in}
\caption{Inter-contact time frequencies between a taxi, the source (top) and the destination (bottom).} 
\label{fig:histfit_intContact_of_taxi1}\vspace{-0.2in}
\end{figure}

We collect  inter-contact times of $16$~taxis with the  source and the destination.
In  Fig.~\ref{fig:histfit_intContact_of_taxi1}, we illustrate the  density function of the contact time for a taxi with the source and the destination, which can be seen to be non-exponential. Table \ref{tab:interContactTimesTaxis} shows the contact rates between each taxi and the source (resp.  the destination), denoted by $\lambda_i$ (resp.  $\mu_i$). 

During the simulations, the source generates a message every $24$ hours. As in Section~\ref{ssec:robust}, the source computes the reward to be promised to  relays  assuming an exponentially distributed  inter-contact time distribution.



\begin{table}[h]
\begin{center}
\vspace{-0.15in}
\caption{Contact rates of the taxis with the source and destination.}
\label{tab:interContactTimesTaxis}
\begin{tabular}{|c|c|c||c|c|c|}
\hline
 Relay &  $\lambda$  &  $\mu$ &  Relay &  $\lambda$  &  $\mu$ \\
\hline
 t$_1$ &  0.0613  &  0.0298 & t$_6$ & 0.0731   & 0.0445   \\ 
\hline
 t$_2$ &  0.0423  &  0.0345 & t$_7$ &  0.0452  & 0.0322 \\
\hline
 t$_3$ &  0.0616  &  0.0382 & t$_8$ &  0.0691  &  0.0513  \\
\hline
 t$_4$ &  0.0340  &  0.0370 & t$_9$ &  0.0596  &  0.0309 \\
\hline
 t$_5$ &  0.0842  &  0.0510 & t$_{10}$ &  0.1095  &  0.0252 \\
\hline
\end{tabular}
\end{center}
\vspace{-0.2in}
\end{table}

\begin{figure}
\begin{center}
\includegraphics[width=\columnwidth]{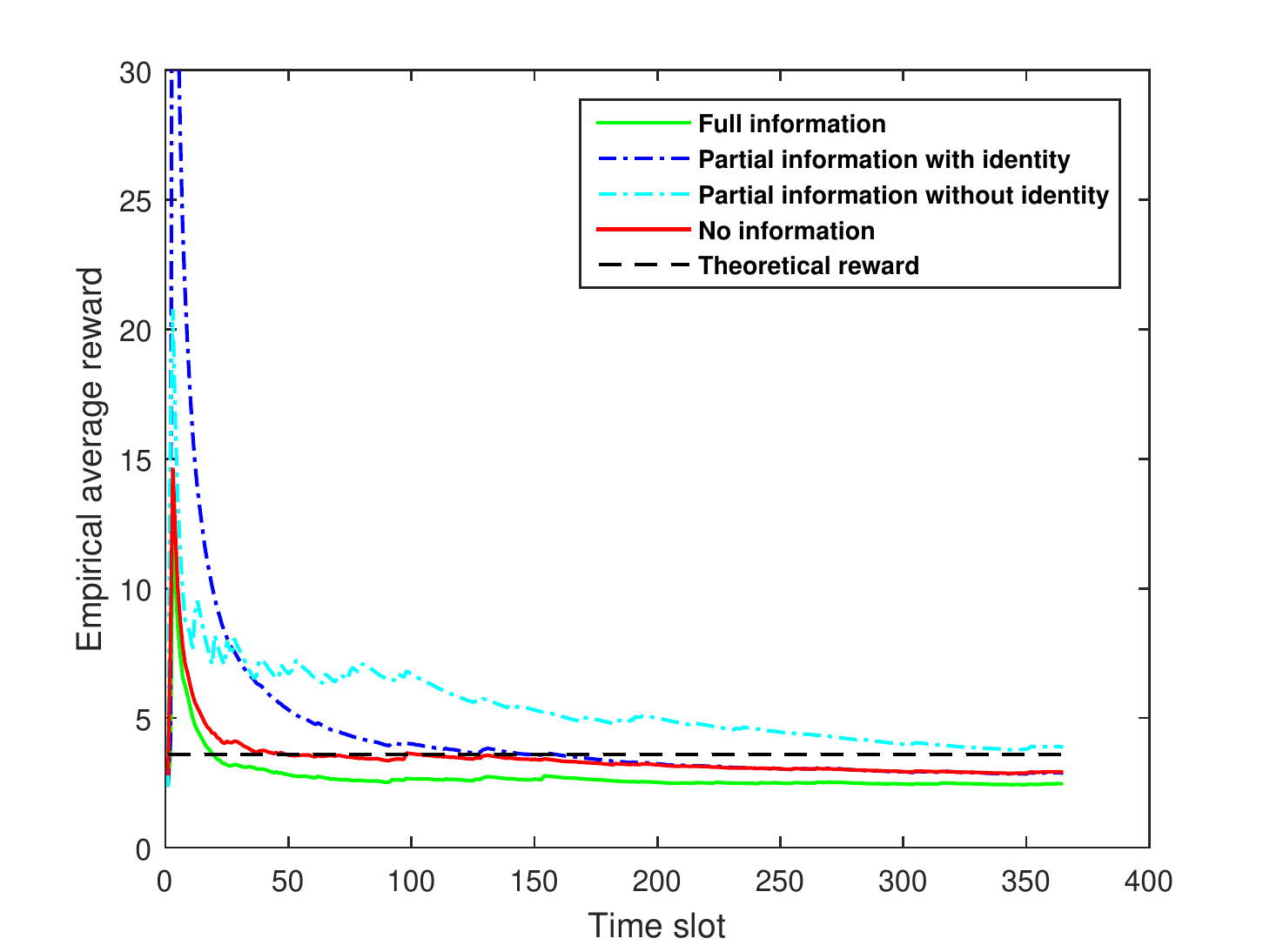}
\end{center}
\vspace{-0.2in}
\caption{Empirical average reward under  four information settings vs the theoretical reward With $N=10$} 
\vspace{-0.2in}
\label{fig:10_relays}
\end{figure}

\begin{figure}
\begin{center}
\includegraphics[width=\columnwidth]{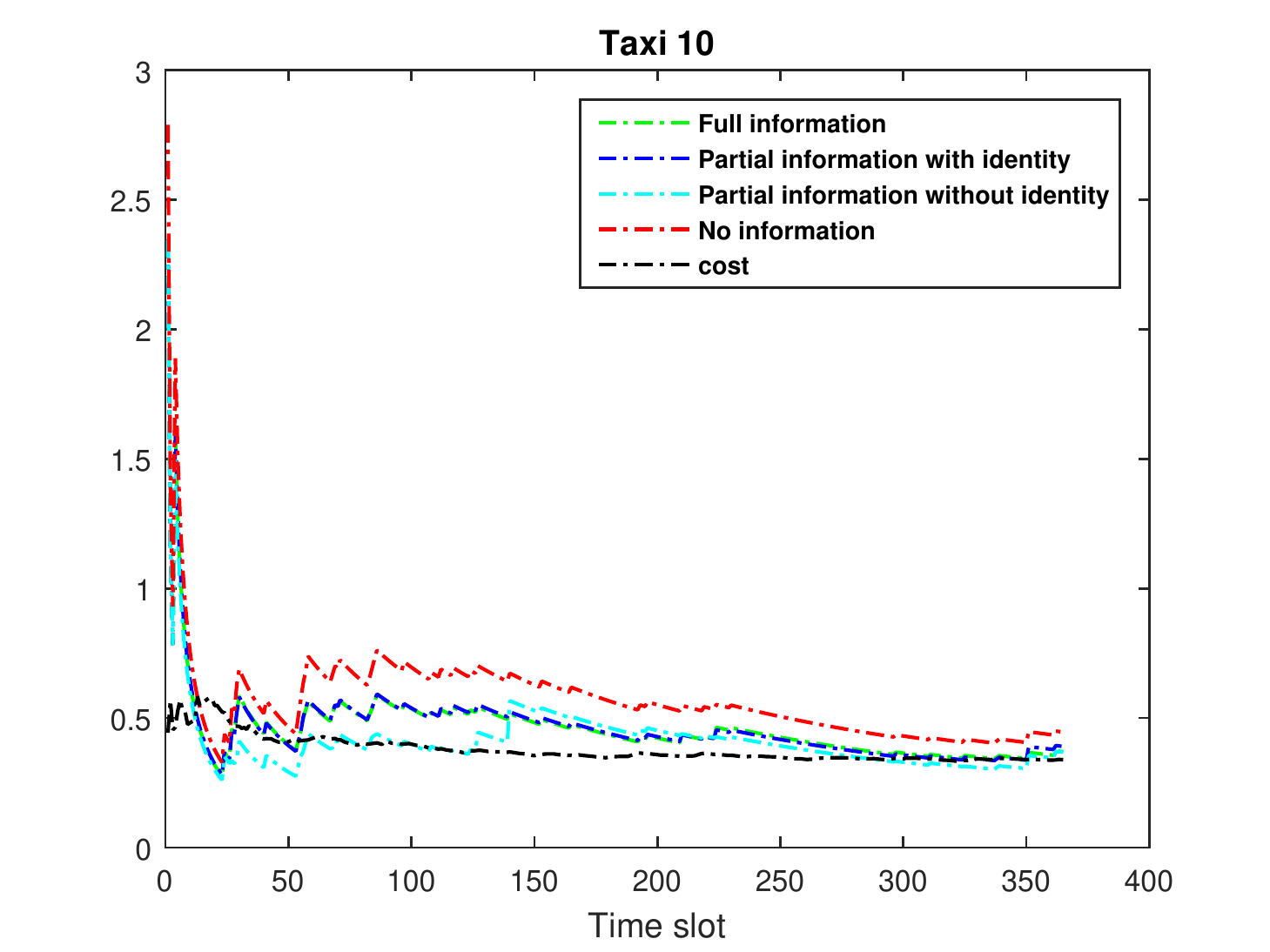}
\end{center}
\vspace{-0.2in}
\caption{Empirical average  reward and energy cost for a relay under four  settings, under static strategies for $N=10$} 
\label{fig:16_relays} \vspace{-0.2in}
\end{figure}

Figure~\ref{fig:10_relays} shows the  empirical average reward paid by the source as a function  time for $10$  taxis and for the four information settings.  The curves  clearly shows the close agreement between the analytical and simulation  results.  Recall that  the expected reward proposed by the source to a relay was  computed analytically under the exponential approximation for   inter-contact times, whereas the  simulation results were obtained using the empirical inter-contact time distribution. Figure~\ref{fig:16_relays}  shows the empirical average  reward as well the expected cost as a function of the time slot.  It is interesting to note that the empirical average reward earned by the taxis, in the four information setting, converges to the theoretical reward calculated using the proposed  analytical model.

\section{Discussion of Assumptions and Limitations}
\label{sec:discuss}
In this section we discuss the main assumptions that were adopted to yield a tractable model. 

\paragraph{Information setting}  We  assume that mobile relays   know the inter-contact time distribution between every node and  the source as well as the destination.   This assumption may  be restrictive in some settings. Nonetheless, with  growing mobile access to the Internet 
 and cloud-based services, we envision 
 a shared knowledge  base maintained by the nodes.   Such knowledge base  makes our incentive reward mechanism realistic in practical settings such as 
 those related to smart cities.  
The analysis of real-world traces indicates that our results are robust against the distribution of inter-contact times. 
Thus, making use of the exponential assumption for inter-contact times, knowledge of contact rates suffices to obtain a good assessment of costs and rewards. 

%

\paragraph{Mobility patterns} A key challenge in developing our results has been to make general assumptions about the mobility of DTN nodes. In particular, the properties derived from the proposed model hold under any set of time-invariant heterogeneous  mobility patterns.  
 It is well known that human mobility patterns change during a day. 
 Future work consists in  accounting for   time-variant mobility  processes, and extending our results to consider time-of-the-day effects.

\paragraph{Buffer management}  In this paper we presented preliminary results on  the impact of time to live (TTL) counters on the delivery probability and relay memory usage, accounting for a single message to be transmitted from a source to its destination.   Buffer management using  TTL caches 
 allows us to  trade between   delivery probability, memory footprint  and  congestion in the network.  
 A detailed analysis of the impacts of the TTL parameters on system performance, as well as the study of the interplay between the delivery of multiple contents, are left as subject for future work.


\section{Conclusion}
\label{sec:conclusion}
In this work, we investigated 
 incentive mechanisms 
 for  heterogeneous mobile networks. We considered a  source node  working under one of four information sharing structures 
 and showed that the expected reward to be paid by the source to relays  remains the same irrespectively  of the information it conveys. Finally,  sensitivity analysis performed using  extensive simulations   indicated  that the average reward paid by the source  under different  information sharing assumptions, in realistic settings, closely matches   the expected reward estimated by our model.

\bibliographystyle{plain}
\bibliography{mybibfile}

\appendix
\section{APPENDIX A}
\subsection{Proof of proposition \ref{propo:exponential}}
Assume that the inter-contact times between a relay $i=1,..,N$ and the source (resp. the destination) fallow an exponential distribution with rate $\lambda_i$ (resp. $\mu_i$).\\
\subsubsection{\textbf{Full information setting }} The source gives information about when (the instant time) and who (identity) is encountered before. Thus, we know the vector of the contact times $\textbf{s} = (s_1, ..,s_n)$ and $\overrightarrow{\ell}$  where $\ell(i)$ denotes the relay who encounter the source at $s_i, i=1, ..,n$.\\
The probability that none of the relays who got the message before time $s_n$ have not yet deliver it to the destination is $$\prod_{k=1}^{n-1} e^{-\mu_{\ell(k)}(s_n-s_k)}$$.

Next, we calculate the probability of a relay to be the first to encounter the destination. For example, Let $X_1, .., X_n$ be an exponential random variables with respective parameters $\lambda_1, .., \lambda_n$, the probability that the minimum is $X_i$ is $\lambda_i / \sum_{j=1}^{n}\lambda_j$. Thus, the probability that the relay $\ell(n)$ is the first to meet the destination among the relays $1, ..,n-1$ and before that the other relays, i.e. relays that are not yet meet the source, get the message from the source is, 
\begin{equation}
\frac{\mu_{\ell(n)}}{\sum_{l=n+1}^{N} \lambda_{\ell(l)} + \sum_{l=1}^{n} \mu_{\ell(l)}} \nonumber
\end{equation}

For each time interval $(s_i, s_{i+1}], i>n$ the probability that a relay $i$, who have not yet the message, get the message from the source before that the others who have the message deliver it to the destination is: {\footnotesize $$\prod_{k=n+1}^{i} \sum_{j=k}^{N} \frac{\lambda_{\ell(j)}}{\sum_{l=k}^{N} \lambda_{\ell(l)} + \sum_{l=1}^{k-1} \mu_{\ell(l)}}$$}.

Hence, the probability that the relay $\ell(n)$ is the first to deliver the message to the destination in time interval $(s_i, s_{i+1}], n \leq i < N$ is, 
\begin{multline}
\frac{\mu_{\ell(n)}}{\sum_{l=i+1}^{N} \lambda_{\ell(l)} + \sum_{l=1}^{i} \mu_{\ell(l)}} \\
\times \prod_{k=n+1}^{i} \sum_{j=k}^{N} \frac{\lambda_{\ell(j)}}{\sum_{l=k}^{N} \lambda_{\ell(l)} + \sum_{l=1}^{k-1} \mu_{\ell(l)}} \nonumber
\end{multline}

Thus, the probability of the relay $\ell(n)$ to be the first to meet the destination after the instant meeting time $s_n$ is,
\begin{multline}
\label{eq:fullInformationFirstTomeetDestination}
\sum_{i=n}^{N} \frac{\mu_{\ell(n)}}{\sum_{l=i+1}^{N} \lambda_{\ell(l)} + \sum_{l=1}^{i} \mu_{\ell(l)}} \\
\times \prod_{k=n+1}^{i} \sum_{j=k}^{N} \frac{\lambda_{\ell(j)}}{\sum_{l=k}^{N} \lambda_{\ell(l)} + \sum_{l=1}^{k-1} \mu_{\ell(l)}} 
\end{multline}

\subsubsection{\textbf{Partial information setting with identity information}}
The source informs the relay about who got the message before but not about the instant time when they had the message. We assume that the source informs the relay about the identity of the encountered relays. Next we are going to calculate, $p_n^{(P^+)}(s_n)$, the probability of success as estimated by the relay $n$ who met the source at $s_n$.

\begin{proof}
$p_n^{(P^+)}(s_n)$ is $(1)$ the probability that none of the relays, who had the message before, have not yet met the destination and $(2)$ the probability that the relay $n$ is the first to meet the destination among all relays who have the message.

$(1)$ The probability that a relay $i=1, ..,n-1$ who have the message before $s_n$ does not yet meet the destination can be expressed as,
\begin{multline}
\int_{0}^{s_n} \frac{\lambda_i e^{-\lambda_i s} e^{-\mu_i(s_n-s)}}{1-e^{-\lambda_i s_n}} \nonumber \\
          = \begin{cases} \frac{\lambda_i}{\lambda_i-\mu_i} \frac{e^{-\mu_i s_n} - e^{-\lambda_i s_n}}{1 - e^{-\lambda_i s_n}}, & \mbox{if } \lambda_i \neq \mu_i, \\
\lambda_i s_n \frac{e^{-\lambda_i s_n}}{1-e^{-\lambda_i s_n}}, & \textrm{otherwise}
\end{cases}
\end{multline}

Thus, the probability that none of the relays $1, .., n-1$ that received the message before $s_n$ did not deliver it to the destination is, 
\begin{align}
\begin{cases}
\prod_{i=1}^{n-1} \frac{\lambda_{i}(e^{-\mu_{i}s_n} - e^{-\lambda_{i}s_n})}{(\lambda_{i}-\mu_{i})(1-e^{-\lambda_{i}s_n})}, & \textrm{if } \lambda_i \neq \mu_i, \vspace{0.5cm}\\
\prod_{i=1}^{n-1}  \lambda_i s_n \frac{e^{-\lambda_i s_n}}{1-e^{-\lambda_i s_n}}, & \textrm{otherwise} 
\end{cases}
\end{align}

$(2)$ The probability that the relay $n$ is the first to deliver the message to the destination is calculated as in full information setting, equation (\ref{eq:fullInformationFirstTomeetDestination}).

Hence, 
\begin{multline}
p_n^{(P^+)}(s_n) = \prod_{i=1}^{n-1}\lambda_{i}\psi(i,n) \times \\
\sum_{i=n}^{N}  \prod_{k=n+1}^{i} \sum_{j=k}^{N}\frac{\lambda_j}{\sum_{l=k}^N \lambda_l + \sum_{l=1}^{k-1} \mu_l}  \frac{\mu_n}{\sum_{l=i+1}^N \lambda_{l} + \sum_{l=1}^i \mu_l } \nonumber
\end{multline}
where
\begin{equation}
\psi(i,n)=\left\{
\begin{array}{ll}
\frac{(e^{-\mu_{i}s_n} - e^{-\lambda_{i}s_n})}{(\lambda_{i}-\mu_{i})(1-e^{-\lambda_{i}s_n})}, & \textrm{if } \lambda_i \neq \mu_i, \\
  \frac{s_n e^{-\lambda_i s_n}}{1-e^{-\lambda_i s_n}}, & \textrm{otherwise} \nonumber
 \end{array}
 \right.
 \end{equation}
\end{proof}

\subsubsection{\textbf{Partial information setting without identity information}}
The source informs the encountered relay only about the number of relays that have the message. The relay is informed only about how many relays got the message before. We are going to calculate the probability of success as estimated by a relay $n$ which met the source at the instant time $s_n$.

\begin{proof}
The probability that the relay $n$ is the first to deliver the message to the destination is: $(1)$ the probability that the relay who got the message before have not yet delivered it to the destination and $(2)$ the probability to be the first to meet the destination among all relays who have the message.\\

$(1)$ As the relay does not know the identity of relays who got the message before, thus the probability that the message has not reached the destination is: the sum of all possible ordering of $n-1$ element from $N-1$ that the $n-1$ relays met the source and they have not delivered the message to the destination yet times the probability that the other relays (i.e., $(N-1)-(n-1)$) have not met the source. this probability is calculated as follows

\begin{equation}
\sum_{i=1}^{C_{N-1}^{n-1}} \prod_{j=1}^{n-1} \lambda_{\ell_i(j)}\psi(\ell_i(j),n) \prod_{k=n+1}^{N} e^{-\lambda_{\ell_i(k)}s_n}
\end{equation}

$(2)$ The probability that the relay $n$ is the first to deliver the message to the destination is calculated as in full information setting, equation (\ref{eq:fullInformationFirstTomeetDestination}).

Hence, the probability that the relay $n$ is the first one who delivers the message to the destination knowing that $n-1$ they have already the message is:
\begin{multline} 
p_n^{(P^-)}(s_n) = \sum_{i=1}^{C_{N-1}^{n-1}} \prod_{j=1}^{n-1}\lambda_{\ell_i(j)}\psi(\ell_i(j),n) \prod_{k=n+1}^{N} e^{-\lambda_{\ell_i(k)}s_n} \times \\
\sum_{i=n}^{N}  \prod_{k=n+1}^{i} \sum_{j=k}^{N}\frac{\lambda_j}{\sum_{l=k}^N \lambda_l + \sum_{l=1}^{k-1} \mu_l}  \frac{\mu_n}{\sum_{l=i+1}^N \lambda_{l} + \sum_{l=1}^i \mu_l } \nonumber
\end{multline}
where
\begin{equation}
\psi(\ell_i(j),n)=\left\{
\begin{array}{ll}
\frac{(e^{-\mu_{\ell_i(j)}s_n} - e^{-\lambda_{\ell_i(j)}s_n})}{(\lambda_{\ell_i(j)}-\mu_{\ell_i(j)})(1-e^{-\lambda_{\ell_i(j)}s_n})}, & \hspace*{-7mm} \textrm{if } \lambda_{\ell_i(j)} \neq \mu_{\ell_i(j)}, \\
 s_n {e^{-\lambda_{\ell_i(j)} s_n}}/({1-e^{-\lambda_{\ell_i(j)} s_n}}), & \textrm{otherwise} \nonumber
 \end{array}
 \right.
 \end{equation}

\end{proof} 

\subsubsection{\textbf{No information setting}}
The source does not give any information to the relay. The relay only knows the instant meeting time $s$ with the source. 

\begin{proof}
Given $s$ the instant meeting time of a relay $n$, $1 \leq n \leq N$ with the source 
and note by $\lambda_i$ (resp. $\mu_i$) the meeting rate of a relay $i, i \leq N$ with the source (resp. the destination). The probability that $m-1$ relays from $N-1$ have already met the source before time $s$ is $$ {C_{N-1}^{m-1}} \prod_{j=1}^{m-1} (1-e^{-\lambda_{\ell_i(j)}}) \prod_{k=m+1}^{N} e^{-\lambda_{\ell_i(k)}}$$

The probability of success of the relay $i$ if he knows how many nodes have already the message is $p_i^{(P^+)}(s)$. Hence, the probability of success of a relay $n$, $ 1 \leq n \leq N$ if it does not have any information is:
$$p_n^{(N)}(s) = \sum_{m=1}^{N} \sum_{i=1}^{C_{N-1}^{m-1}} \prod_{j=1}^{m-1} (1-e^{-\lambda_{\ell_i(j)}}) \prod_{k=m+1}^{N} e^{-\lambda_{\ell_i(k)}} p_m^{(P^+)}(s) $$
\end{proof}

\section{APPENDIX B}
\subsection{Proof of lemma \ref{lemma:ExpectedRewardPaidByTheSource}} 
\begin{proof}
The expected reward paid by the source under setting $(k)$ is
\begin{align}
& R^{(k)} \nonumber  \\
&=  c_d + \int_{\tiny{s_{1}=0:s_N=s_{N-1}}}^{\infty} \sum_{\bmell_m \in L} R^{(k)}(\bss,\bmell_m) f(\bss, \bmell_m) ds_N \ldots ds_1 \nonumber  \\
& = c_d + \int_{s_{1}=0:s_N=s_{N-1}}^{\infty} \sum_{\bmell_m \in L}  \sum_{i=1}^{N} \mathcal{C}_{\bmell_m{(i)}} \frac{ p_i(\bss, \bmell_m)}{ p_i^{(k)}(\bss, \bmell_m)}  f(\bss, \bmell_m) ds_{N:1} \nonumber  \\
&=  c_d + \sum_{\bmell_m \in L} \sum_{i=1}^{N} \mathcal{C}_{\bmell_m{(i)}}  \int_{s_{1} =0:s_N=s_{N-1}}^{\infty} \frac{ p_i(\bss, \bmell_m)}{ p_i^{(k)}(\bss, \bmell_m)}  f(\bss, \bmell_m) ds_{N:1} \nonumber \\
\end{align}
if we set
\begin{equation}
\varphi_i(\bss,\bmell_m) =  \int_{s_{i+1}=s_i:s_N=s_{N-1}}^{\infty} \hspace{-10mm} p_i(\bss, \bmell_m) f(\bss_{i+1:N},\bmell_m) ds_{{N:i+1}} 
\end{equation}

we get the expression of Lemma \ref{lemma:ExpectedRewardPaidByTheSource}:
\begin{align}
& R^{(k)} =\nonumber  \\
& c_d + \sum_{\bmell_m \in L} \sum_{i=1}^{N} \mathcal{C}_{\bmell_m{(i)}}  \int_{s_{1}=0:s_i=s_{i-1}}^{\infty} \frac{  f(\bss_{1:i}, \bmell_m)}{ p_i^{(k)}(\bss, \bmell_m)} \varphi_i(\bss,\bmell_m) ds_{{i:1}} \nonumber \\
\end{align}
\end{proof}
\subsection{Proof of Theorem \ref{theorem1}}
\begin{proof}
We calculate the expected reward to be paid by the source in all information settings.
\subsubsection{Full Information Setting}
We relate $p_{j}^{(F)}(\bss, \bmell)$, The probability of success for the the $j$-th relay to meet the source under the full information setting, to $p_{j}(\bss, \bmell)$ by removing the dependency of  $p_{j}(\bss, \bmell)$ on $s_{j+1}, s_{j+2}, \ldots, s_N$. We get 

\begin{small}
\begin{align}
& {p}_{j}^{(F)}(\bss, \bmell) \nonumber \\
& = \hspace{-5mm} \sum_{\stackrel{\bmell_m\in L}{\bmell_m(1:j)=\bmell(1:j)}}\int_{s_{j+1}=s_j:s_N=s_{N-1}}^{\infty} p_{j}(\bss, \bmell_m) {f}_{j+1:N}(\bss,\bmell_m) ds_{N:j+1}  \nonumber \\
& =\frac{\sum_{\stackrel{\bmell_m\in L}{\bmell_m(1:j)=\bmell(1:j)}}  \int_{s_j:s_{N-1}}^{\infty} p_{j}(\bss, \bmell_m) { \prod_{i=j+1}^N f^{\source}_{i}(s_i,\bmell_m)} ds_{N:j+1}}{ \sum_{\stackrel{\bmell_m\in L}{\bmell_m(1:j)=\bmell(1:j)}} \int_{s_j:s_{N-1}}^{\infty}{ \prod_{i=j+1}^N f^{\source}_{i}(s_i,\bmell_m)} ds_{N:j+1}} \label{eq:oldapproach2}
\end{align}
\end{small}

Let define $g_i(s_i, \bmell_m)$  as follows
\begin{small}
\begin{eqnarray*}
 g_i(s_i, \bmell)= \hspace{-7mm} \sum_{\stackrel{\bmell_m\in L}{\bmell_m(1:j)=\bmell(1:j)}}  { \int_{s_{i+1}=s_i:s_N=s_{N-1}}^{\infty}{ \prod_{j=i+1}^N f^{\source}_{j}(s_j,\bmell_m)} ds_{N:i+1}} \label{eq:fullg}
\label{eq:convenient}
\end{eqnarray*}
\end{small}
Thus 
\begin{equation}
p_i^{(F)}(\bss,\bmell)=  \frac{ \sum_{\stackrel{\bmell_m\in L}{\bmell_m(1:j)=\bmell(1:j)}}\varphi_i(\bss,\bmell)}{g_i(s_i, \bmell)}  \label{eq:pipfull}
\end{equation}

Now we are going to calculate the average reward paid by the source taking into consideration all the possible scenarios for sending a message using the full information setting. From Lemma \ref{lemma:ExpectedRewardPaidByTheSource} we have:
\begin{align}
& R^{(F)} - c_d \nonumber \\
&= \hspace{-2mm} \sum_{m=1}^{N!} \sum_{i=1}^{N} \mathcal{C}_{\bmell_m{(i)}} \hspace{-2mm} \int_{s_{1}=0:s_i=s_{i-1}}^{\infty} \hspace{-14mm} {  f(\bss_{1:i}, \bmell_m)}  {g}_i(s_i, \bmell_m) \frac{\varphi_i(\bss,\bmell)}{\sum \varphi_i(\bss,\bmell)} ds_{{i:1}} \nonumber \\ 
&= \sum_{i=1}^{N} \mathcal{C}_{r_i}\Psi_i(\bss)\label{eq:fullrk4} 
\end{align}

where $\Psi_i(\bss) =$
\begin{align}
 \sum_{j=1}^N \hspace{-2mm} \sum_{\stackrel{\bmell_m\in L: }{\bmell_m(j)=r_i} } \hspace{-2mm} \int_{s_{1}=0:s_j=s_{j-1}}^{\infty} \hspace{-14mm} {  f(\bss_{1:j}, \bmell_m)}  {g}_j(s_j, \bmell_m) \frac{\varphi_j(\bss,\bmell)}{\sum \varphi_j(\bss,\bmell)} ds_{{j:1}}  \label{eq:fullrk2} 
\end{align}

Next, we are going to show that $\Psi_i(\bss)=1$. 
\begin{eqnarray*}
\Psi_i(\bss,\bmell) &=& \hspace{-2mm}  \sum_{j=1}^N \sum_{\stackrel{\bmell_m(1:j)\in L_j: }{\bmell_m(j)=r_i} } \sum_{\stackrel{\tilde{\bmell}_m\in L: }{\tilde{\bmell}_m(1:j) = \bmell_m(1:j)} } \\
 & & \int_{s_{1}=0:s_j=s_{j-1}}^{\infty} \hspace{-10mm} {  f(\bss_{1:j})}  {g}_j(s_j, \bmell_m) \frac{\varphi_j(\bss,\bmell_m)}{\sum \varphi_j(\bss,\bmell)} ds_{{j:1}}  \label{eq:fullrk3} \\
&=& \hspace{-2mm} \sum_{j=1}^N \sum_{\stackrel{\forall \bmell_m(1:j): }{\bmell_m(j)=r_i} }  \int_{s_{1}=0:s_j=s_{j-1}}^{\infty} \hspace{-13mm}  {  f(\bss_{1:j}, \bmell_m)}  {g}_j(s_j, \bmell_m)  ds_{{j:1}}  \\
&=& \hspace{-2mm} \sum_{j=1}^N \sum_{\stackrel{\forall \bmell_m(1:j): }{\bmell_m(j)=r_i} } \int_{s_{1}=0:s_j=s_{j-1}}^{\infty} \hspace{-13mm}  {  f(\bss_{1:j}, \bmell_m)}  {g}_j(s_j, \bmell_m)  ds_{{j:1}}  \\
&=& \hspace{-2mm} \sum_{m=1}^{N!} \sum_{j=1}^N  1_{\bmell_j(m)=r_i} \int_{s_{1}=0:s_N=s_{N-1}}^{\infty} \hspace{-12mm}  {  f(\bss_{1:N}, \bmell_m)}   ds_{{N:1}}  \label{eq:fullrk6}  \\
&=& \hspace{-2mm} \sum_{m=1}^{N!}  \int_{s_{1}=0:s_N=s_{N-1}}^{\infty} \hspace{-8mm} {  f(\bss_{1:N}, \bmell_m)}   ds_{{N:1}}  \label{eq:fullrk7}  \\
&=&1
\end{eqnarray*}
Hence, the expected reward to be paid by the source under the full information setting is: 
\begin{equation}
R^{(F)}= c_d + \sum_{i=1}^{N} \mathcal{C}_{r_i}
\end{equation}

\subsubsection{Partial Information setting}
In the partial information setting the source informs the relay about the set of encountered relays (i.e. the relays who had already the message),  but it does not divulge the relays infection time. We distinguish two cases: The source can either 1) divulge the number and the identity of relays who met the source before, let denote this case by $P^{+}$. Or 2) divulge just the number of relays who met the source i.e. the relay will not know the identity of relays, let denote this case by $P^{-}$.
Next we will calculate the expected reward to be paid by the source in the two cases of the partial information setting.

\textbf{Partial Information with identity of relays for previous contacts:} \leavevmode\
From Lemma \ref{lemma:ExpectedRewardPaidByTheSource}, we have:
\begin{align}
& R^{(P^{+})} - c_d = \nonumber \\
& \sum_{m=1}^{N!} \sum_{i=1}^{N} \mathcal{C}_{\bmell_m{(i)}}  \int_{s_{1}=0:s_i=s_{i-1}}^{\infty} \frac{  f(\bss_{1:i}, \bmell_m)}{ p_i^{(P^{+})}(\bss, \bmell_m)} \varphi_i(\bss,\bmell_m) ds_{{i:1}} 
\end{align}


Note that $p_i^{(P^{+})}$ depends on $\bss$ only through $s_i$.  Therefore, we replace the order
of the integrals in the expression above, and write
\begin{align}
 &R^{(P^{+})} =c_d + \sum_{m=1}^{N!} \sum_{i=1}^{N} \mathcal{C}_{\bmell_m{(i)}}  \int_{s_i=0}^\infty \frac{f(\bss_i,\bmell_m)}{ p_i^{(P^{+})}(\bss, \bmell_m)}  \Phi_i(\bss,\bmell_m)ds_i   \label{eq:diffpartial}
\end{align}
where
\begin{align} \label{eq:Phiibss}
&\Phi_i(\bss,\bmell)  \nonumber\\
& = \int_{s_1=0:s_{i-1}=s_{i-2}}^{s_i:s_i} \int_{s_{i+1}=s_i:s_N=s_{N-1}}^{\infty} \hspace{-10mm} p_i(\bss,\bmell) \prod_{\stackrel{j=1}{j \neq i}}^{N} f_j(s_j,\bmell) ds_{-i} \nonumber \\
& = \int_{s_{i-1}=0:s_{1}=0}^{s_i:s_2} \int_{s_{i+1}=s_i:s_N=s_{N-1}}^{\infty} \hspace{-10mm}  p_i(\bss,\bmell)\prod_{\stackrel{j=1}{j \neq i}}^{N} f_j(s_j,\bmell) ds_{-i} \nonumber \\
&= {\int_{s_{i-1}=0:s_{1}=0}^{s_i:s_2} {  f(\bss_{1:i-1}, \bmell)}}   \varphi_i(\bss,\bmell) ds_{{1:i-1}} \nonumber \\
& = \overline{g}_i(s_i,\bmell) \int_{s_{i-1}=0:s_1=0}^{s_i:s_2} \hspace{-6mm}  \hat{p}_i^{(F)}(\bss,\bmell) \prod_{j=1}^{i-1} f_j(s_j,\bmell) ds_{1:i-1}  
\end{align}

Next, we derive a simple expression for $p_{i}^{(P^{+})}(\bss, \bmell)$ as a function of 
$p_{i}^{(F)}(\bss, \bmell)$,   
\begin{align}
&{p}_{i}^{(P^{+})}(\bss, \bmell) \hspace{-3mm} \nonumber\\
&= \int_{s_{1}=0:s_{i-1}=s_{i-2}}^{s_i:s_i} p^{(F)}_{i}(\bss, \bmell)f_{1:i-1}(\bss,\bmell) ds_{i-1:1} \nonumber \\
&=\frac{ \int_{s_{1}=0:s_{i-1}=s_{i-2}}^{s_i:s_i} p^{(F)}_{i}(\bss, \bmell) { \prod_{j=1}^{i-1} f^{\source}_{j}(s_j,\bmell)} ds_{i-1:1}}{ \int_{s_{1}=0:s_{i-1}=s_{i-2}}^{s_i:s_i}{ \prod_{j=1}^{i-1} f^{\source}_{j}(s_j,\bmell)} ds_{i-1:1}} \nonumber \\
&= \frac{ \int_{s_{i-1}=0:s_{1}=0}^{s_i:s_2} p^{(F)}_{i}(\bss, \bmell) { \prod_{j=1}^{i-1} f^{\source}_{j}(s_j,\bmell)} ds_{1:i-1}}{   \int_{s_{i-1}=0:s_{1}=0}^{s_i:s_2} { \prod_{j=1}^{i-1} f^{\source}_{j}(s_j,\bmell)} ds_{1:i-1} } 
\end{align}

We have $p_{i}^{(P^{+})}(\bss, \bmell)$ depends on $\bmell$.  
This is because we are assuming that the $i$-th node in
 $\bmell$
knows the identities of the previous $i-1$ nodes encountered by the source.
Therefore, $p_{i}^{(P^{+})}(\bss, \bmell)$ depends on $\bmell$ through its first $i$ entries.

Let
\begin{equation}
 h_i(s_i,\bmell_m)= \int_{s_{i-1}=0:s_{1}=0}^{s_i:s_2} { \prod_{j=1}^{i-1} f^{\source}_{j}(s_j,\bmell_m)} ds_{1:i-1} 
\end{equation}

Note that whereas the function $h$ integrates  over encounters that occurred previous to $s_i$ (backward), the function $g$ integrates
over future encounters (forward).   Then, $p_i^{(P^{+})}(\bss,\bmell)$ can be written as a function of $g$ and $h$ as follows, 
\begin{align}
&p_{i}^{(P^{+})}(\bss, \bmell) =  \frac{1}{\overline{g}_i(s_i,\bmell)  h_i(s_i,\bmell) }\sum_{\stackrel{\forall \bmell_m:}{\bmell_m(1:i)=\bmell(1:i)}} \Phi_i(\bss,\bmell_m) \label{eq:pip}
\end{align} 
Note that~\eqref{eq:pip} is analogous to~\eqref{eq:pipfull}.   
Replacing \eqref{eq:pip} into~\eqref{eq:diffpartial} we get, 
\begin{align}
& R^{(P^{+})} - c_d \nonumber \\
& =  \sum_{m=1}^{N!} \sum_{i=1}^{N} \mathcal{C}_{\bmell_m{(i)}}  \int_{s_i=0}^\infty \frac{f(\bss_i,\bmell_m)}{ p_i^{(P^{+})}(\bss, \bmell_m)}  \Phi_i(\bss,\bmell_m)ds_i  \nonumber \\
& = \sum_{i=1}^{N} \mathcal{C}_{r_i} \sum_{j=1}^N \sum_{\stackrel{m=1}{\bmell_m(j)=r_i}}^{N!}   \int_{s_j=0}^\infty \frac{f(\bss_j,\bmell_m)}{ p_j^{(P^{+})}(\bss, \bmell_m)}  \Phi_j(\bss,\bmell_m)ds_j     \nonumber \\
& = \sum_{i=1}^{N} \mathcal{C}_{r_i} \Xi_i(\bss,\bmell)
\end{align}
where
\begin{align}
  \Xi_i(\bss,\bmell) = \sum_{j=1}^N \sum_{\stackrel{m=1}{\bmell_m(j)=r_i}}^{N!}   \int_{s_j=0}^\infty \frac{f(\bss_j,\bmell_m)}{ p_j^{(P^{+})}(\bss, \bmell_m)}  \Phi_j(\bss,\bmell_m)ds_j    
\end{align}

Next, we are going to show that $ \Xi_i(\bss,\bmell)=1$.
\begin{align}
& \Xi_i(\bss,\bmell) \nonumber \\
&=  \sum_{j=1}^N \sum_{\stackrel{m=1}{\bmell_m(j)=r_i}}^{N!}   \int_{s_j=0}^\infty \frac{f(\bss_j,\bmell_m)}{ p_j^{(P_{+})}(\bss, \bmell_m)}  \Phi_j(\bss,\bmell_m)ds_j  \nonumber \\
& = \sum_{j=1}^N \sum_{\stackrel{\forall \bmell_m(1:j): }{\bmell_m(j)=r_i} } \sum_{\stackrel{\forall \tilde{\bmell}_m: }{\tilde{\bmell}_m(1:j) = \bmell_m(1:j)} } \hspace{-8mm} 
\int_{s_j=0}^\infty \frac{f(\bss_j,\bmell_m)}{ p_j^{(P_{+})}(\bss, \bmell_m)}  \Phi_j(\bss,\tilde{\bmell}_m)ds_j \nonumber \\
& =  \sum_{j=1}^N \sum_{\stackrel{\forall \bmell_m(1:j): }{\bmell_m(j)=r_i} } \int_{s_j=0}^\infty \frac{f(\bss_j,\bmell_m) \overline{g}_j(s_j,\bmell_m) h_j(s_j,\bmell_m)}{ \sum_{\stackrel{\forall \tilde{\bmell}_m: }{\tilde{\bmell}_m(1:j) = \bmell_m(1:j)} }  \Phi_i(\bss,\tilde{\bmell}_m)} \nonumber \\
& \hspace{35mm} \times \sum_{\stackrel{\forall \tilde{\bmell}_m: }{\tilde{\bmell}_m(1:j) = \bmell_m(1:j)} }  \Phi_j(\bss,\tilde{\bmell}_m)ds_j  \nonumber \\
& =  \sum_{j=1}^N \sum_{\stackrel{\forall \bmell_m: }{\bmell_m(j)=r_i} } \int_{s_j=0}^\infty {f(\bss_j,\bmell_m) {g}_i(s_i,\bmell_m) h_i(s_i,\bmell_m)}  ds_j   \nonumber \\
 &= 1 
\end{align}

Hence, the expected reward paid by the source under the partial information setting with identity information about previous contacts is : 
\begin{equation}
R^{(P^+)} = c_d + \sum_{i=1}^{N} \mathcal{C}_{r_i}
\end{equation}

\textbf{Partial Information Without Identity Information About Previous Contacts:}\leavevmode\
Now consider the partial information setup wherein relays are not informed about the identities of relays that have already encountered the source in the previous contacts. As the arguments are very similar to the ones presented in the previous setup, we focus only on the key steps.

First, we derive $ p_{i}^{(P^+)}(\bss,\bmell)$, which is a function of $i$, $s_i$ and $\bmell(i)$.  
\begin{align}
& p_{i}^{(P^+)}(\bss,\bmell) \nonumber \\
& = \frac{\sum_{\stackrel{\forall \bmell_m:}{\bmell_m(i)=\bmell(i)}} \int_{0:s_{i-1}}^{s_i:s_i} \int_{s_i:s_{N-1}}^{\infty}p_i(\bss,\bmell_m) \prod_{\stackrel{j=1}{j \neq i}}^{N} f_j(s_j,\bmell_m) ds_{-i}}{\sum_{\stackrel{\forall \bmell_m:}{\bmell_m(i)=\bmell(i)}} \int_{s_1=0:s_{i-1}=s_{i-2}}^{s_i:s_i} \int_{s_i:s_{N-1}}^{\infty} \prod_{\stackrel{j=1}{j \neq i}}^{N} f_j(s_j,\bmell_m)    ds_{-i}} \nonumber \\
&= \frac{1}{\overline{h}_i(s_i,\bmell_m) }  \sum_{\stackrel{\forall \bmell_m:}{\bmell_m(i)=\bmell(i)}} \Phi_i(\bss,\bmell_m) \label{eq:partialnoprevious}
\end{align}

where

\begin{eqnarray}
 \overline{h}_i(s_i,\bmell) &=& \hspace{-6mm} \sum_{\stackrel{\forall \bmell_m:}{\bmell_m(i)=\bmell(i)}} \int_{0:s_{i-2}}^{s_i:s_i} \int_{s_i:s_{N-1}}^{\infty} \prod_{\stackrel{j=1}{j \neq i}}^{N} f_j(s_j,\bmell_m) ds_{-i}  \nonumber\\
 \hspace{-8mm} &=&  \hspace{-6mm} \sum_{{\stackrel{\forall \bmell_m(1:i):}{\bmell_m(i)=\bmell(i)}}}  \hspace{-2mm} \int_{0:0}^{s_i:s_2} { \prod_{j=1}^{i-1} f^{\source}_{j}(s_j,\bmell_m)} g_i(s_i,\bmell_m) ds_{1:i-1} 
\end{eqnarray}

Note that~\eqref{eq:partialnoprevious} is analogous to~\eqref{eq:pip}. Considering now  the
 partial information setup without information about previous contacts.

Replacing \eqref{eq:partialnoprevious} into~\eqref{eq:diffpartial}, 

\begin{align}
& R^{(P^{-})} -c_d \nonumber \\
&= \sum_{m=1}^{N!} \sum_{i=1}^{N} \mathcal{C}_{\bmell_m{(i)}}  \int_{s_i=0}^\infty \frac{f(\bss_i,\bmell_m)}{ \overline{p}_i^{(P^{-})}(\bss, \bmell_m)}  \Phi_i(\bss,\bmell_m)ds_i  \nonumber \\
&=\sum_{i=1}^{N} \mathcal{C}_{r_i} \sum_{j=1}^N \sum_{\stackrel{m=1}{\bmell_m(j)=r_i}}^{N!}   \int_{s_j=0}^\infty \frac{f(\bss_j,\bmell_m)}{ \overline{p}_j^{(P^{-})}(\bss, \bmell_m)}  \Phi_j(\bss,\bmell_m)ds_j \nonumber \\
&=\sum_{i=1}^{N} \mathcal{C}_{r_i} \overline{\Xi}_i(\bss,\bmell)
\end{align}

where
\begin{align}
  \overline{\Xi}_i(\bss,\bmell) = \sum_{j=1}^N \sum_{\stackrel{m=1}{\bmell_m(j)=r_i}}^{N!}   \int_{s_j=0}^\infty \frac{f(\bss_j,\bmell_m)}{ \overline{p}_j^{(P^{-})}(\bss, \bmell_m)}  \Phi_j(\bss,\bmell_m)ds_j    
\end{align}
Next, we are going to show that $ \overline{\Xi}_i(\bss,\bmell)=1$.
\begin{align*}
& \overline{\Xi}_i(\bss,\bmell) \\
& = \sum_{j=1}^N \sum_{\stackrel{m=1}{\bmell_m(j)=r_i}}^{N!}   \int_{s_j=0}^\infty \frac{f(\bss_j,\bmell_m)}{ \overline{p}_j^{(P^{-})}(\bss, \bmell_m)}  \Phi_j(\bss,\bmell_m)ds_j  \\
&  = \sum_{j=1}^N    \int_{s_j=0}^\infty \frac{f_{r_i}(s_j)}{ \overline{p}_{j}^{(P^{-})}(s_j,r_{i})}  \sum_{\stackrel{m=1}{\bmell_m(j)=r_i}}^{N!} \Phi_j(\bss,\bmell_m)ds_j  \\
& = \sum_{j=1}^N    \int_{s_j=0}^\infty \frac{f_{r_i}(s_j)\overline{h}_j(s_j,\bmell)}{ \sum_{\stackrel{m=1}{\bmell_m(j)=r_i}}^{N!} \Phi_j(\bss,\bmell_m)} \hspace{-3mm} \sum_{\stackrel{m=1}{\bmell_m(j)=r_i}}^{N!} \hspace{-3mm} \Phi_j(\bss,\bmell_m)ds_j  \\
& = \sum_{j=1}^N    \int_{s_j=0}^\infty {f_{r_i}(s_j)\overline{h}_j(s_j,\bmell)}  ds_j  \\
& = \sum_{m=1}^{N!}   g(\bmell_m)  \\
& = 1
\end{align*}

\subsubsection{No information}
The source doesn't give any information about the previous contacts to the encountered relay. Recall that $p_{r_i}^{(N)}(s)$ is the probability of the relay $r_i$ to be the first to meet the destination, given that it encountered the source at time $s$,  as 
estimated by such node.  
 
We start by deriving an expression for $p_{r_i}^{(N)}(s)$ as a function of $\Phi_i(\bss,\bmell)$; the probability that the $i$-th node, to meet the source under ordering $\bmell$, is the first to meet the destination, as estimated by such node which is given by~\eqref{eq:Phiibss}.  
  As $\Phi_i(\bss,\bmell)$ depends on $\bss$ only through $s_i$, we denote it by
  $\Phi_i(s,\bmell)$.  
  Then,
\begin{align} \label{eq:equival1}
& p_{r_i}^{(N)}(s) =   \sum_{m=1}^{N!}  \Phi_{\bmell_m^{-1}(r_i)}(s,\bmell_m)
\end{align}

Alternatively, $p_{r_i}^{(N)}(s)$ is given as follows,
\begin{align} \label{eq:equival2}
p_{r_i}^{(N)}(s) &=   \sum_{n=1}^{N}  \sum_{\forall \bmell_m: \bmell_m(n)=r_i}  \Phi_{n}(s,\bmell_m)
%
%
\end{align}

To see the equivalence between~\eqref{eq:equival1} and~\eqref{eq:equival2},
we  note that it follows from ~\eqref{eq:equival2} that
\begin{align} \label{eq:probNoInformation}
p_{r_i}^{(N)}(s) &=    \sum_{\forall \bmell_m} \sum_{n=1}^{N} 1_{\bmell_m(n)=r_i} \Phi_{n}(s,\bmell_m) \nonumber\\
&=    \sum_{m=1}^{N!} \sum_{n=1}^{N} 1_{\bmell_m(n)=r_i} \Phi_n(s,\bmell_m) \nonumber\\
& = \sum_{m=1}^{N!}  \Phi_{\bmell_m^{-1}(r_i)}(s,\bmell_m)
\end{align}
where $1_{c}$ denotes an indicator  function, equal to 1 if condition $c$ holds and 0 otherwise.

From~\eqref{eq:averageRewardGenralCase} we have that

\begin{align} 
& R^{(N)} - c_d \nonumber\\
&=  \sum_{m=1}^{N!} \sum_{i=1}^{N} \mathcal{C}_{\bmell_m{(i)}}  \int_{s_{1}=0:s_N=s_{N-1}}^{\infty} \frac{  f(\bss, \bmell_m)p_{i}(\bss, \bmell_m)}{ p_{\bmell_m(i)}^{(N)}(\bss, \bmell_m)}  ds_{{N:1}}  \nonumber \\
&= \sum_{i=1}^{N} \mathcal{C}_{r_i}  \sum_{m=1}^{N!}  \int_{s_{1}=0:s_N=s_{N-1}}^{\infty} \frac{  f(\bss, \bmell_m)p_{\bmell^{-1}_m(r_i)}(\bss, \bmell_m)}{ p_{r_i}^{(N)}(s_j)}  ds_{{N:1}}  \nonumber \\
&= \sum_{i=1}^{N} \mathcal{C}_{r_i}  \sum_{j=1}^N \sum_{\stackrel{\forall \bmell_m:}{\bmell_m(j)=r_i}}  \int_{0:s_{N-1}}^{\infty} \frac{  f(\bss, \bmell_m)p_{j}(\bss, \bmell_m)}{ p_{r_i}^{(N)}(s_j)}  ds_{{N:1}} \nonumber \\
&= \sum_{i=1}^{N} \mathcal{C}_{r_i} \Gamma(i)  \label{eq:beforeeq1}
\end{align}

where
\begin{equation}
\Gamma(i) =\sum_{j=1}^N \sum_{\stackrel{\forall \bmell_m:}{\bmell_m(j)=r_i}}  \int_{s_{1}=0:s_N=s_{N-1}}^{\infty} \hspace{-4mm} \frac{  f(\bss, \bmell_m)p_{j}(\bss, \bmell_m)}{ p_{r_i}^{(N)}(s_j)}  ds_{{N:1}}  
\end{equation}
Next, we are going to show that $\Gamma(i)=1$.
\begin{eqnarray}
\label{eq:equal1}
\Gamma(i) &= & \hspace{-2mm} \sum_{\forall \bmell_m}  \sum_{j=1}^N  1_{\bmell_m(j)=r_i} \int_{0:s_{N-1}}^{\infty} \hspace{-4mm} \frac{  f(\bss, \bmell_m)p_{j}(\bss, \bmell_m)}{ p_{r_i}^{(N)}(s_j)}  ds_{{N:1}}  \nonumber\\
&= & \hspace{-2mm} \sum_{\forall \bmell_m}  \sum_{j=1}^N  1_{\bmell_m(j)=r_i} \int_{s_j=0}^{\infty} \frac{  f_{r_i}(s_j)}{ p_{r_i}^{(N)}(s_j)}  \Phi_j(\bss,\bmell_m) ds_j \nonumber \\
&= & \hspace{-2mm}  \int_{s=0}^{\infty} \frac{  f_{r_i}(s)}{ p_{r_i}^{(N)}(s)}  \sum_{\forall \bmell_m}  \sum_{j=1}^N  1_{\bmell_m(j)=r_i} \Phi_j(\bss,\bmell_m) ds  \nonumber \\
&= & \hspace{-2mm} \int_{s=0}^{\infty} {  f_{r_i}(s)} ds  \nonumber \\
&= & \hspace{-2mm} 1  
\end{eqnarray}

Therefore,~\eqref{eq:beforeeq1} together with~\eqref{eq:equal1} imply that 
\begin{equation}
R^{(N)} =c_d + \sum_{i=1}^{N} \mathcal{C}_{r_i}
\end{equation}
\end{proof}

\end{document}